\documentclass[11pt,a4paper]{article}

\usepackage[utf8]{inputenc}
\usepackage[T1]{fontenc}
\usepackage{lmodern}
\usepackage[margin=1in]{geometry}
\usepackage{amsmath, amssymb, amsthm}
\usepackage{graphicx}
\usepackage{booktabs}
\usepackage{multirow}
\usepackage{enumitem}
\usepackage{hyperref}
\usepackage{xcolor}
\usepackage[round,authoryear]{natbib}
\usepackage{titlesec}
\usepackage{authblk}

\hypersetup{colorlinks=true, citecolor=blue, linkcolor=black, urlcolor=blue}

\title{Bitcoin's Power Law: Weak Structure, Strong Forecasts}
\author[a]{Carlos Baquero\thanks{\texttt{cbm@fe.up.pt}}}
\author[b]{Raquel Menezes\thanks{\texttt{rmenezes@math.uminho.pt}}}
\affil[a]{Faculty of Engineering, University of Porto and INESC TEC, Portugal}
\affil[b]{Centro de Matem\'{a}tica, Universidade do Minho, Braga, Portugal}
\date{\today}

\begin{document}
\maketitle

\begin{abstract}
Bitcoin's price has been described as following a power law (PL)
in time, $P \sim t^{\beta}$ with $\hat\beta \approx 5.7$ over
2010--2026. We test this claim using the
Clauset--Shalizi--Newman protocol applied to Bitcoin's
tail-relevant distributional series, and develop three principled
time-domain adaptations of the protocol. We find that
(i) the distributional power law is rejected on UTXO balances and
daily $|$returns$|$, with lognormal preferred decisively;
(ii) the fitted time-domain exponent varies by nearly a factor of
three across reasonable shifts of the time origin --- it is not
specification-robust in the sense required for a shift-invariant
structural reading; (iii) standard residual diagnostics and
scale-invariance tests proposed in earlier work cannot distinguish
a power law from a multi-component sigmoid stack fit to the same
data; (iv) Bitcoin price stands apart in a cross-asset comparison
spanning Bitcoin on-chain metrics and traditional asset classes:
it is the only series in the nine-series in-sample test where no
single-component growth curve improves on the power law, and the
quarterly $K = 3$ wave-stability bootstrap rejects the PL+AR(1)
null on Bitcoin at $p_{<15\%} = 0.015$ --- a clear cross-asset
separation, although not a Bonferroni-robust rejection; and
(v) walk-forward Diebold--Mariano evaluation against ten
candidates --- including standard time-series baselines (RW with
drift, auto-ARIMA, ETS, local-linear-trend) --- shows the
in-sample winner (multi-sigmoid) is among the worst long-horizon
forecasters, while the simple power law dominates 12--24 month
horizons against every standard baseline at $p < 0.05$, precisely
because it does not commit to specific wave shapes. The
fit--prediction tradeoff is the practical counterpart of the
descriptive findings.
\end{abstract}


\bigskip

\section{Introduction}
\label{sec:intro}

Bitcoin has been continuously traded for roughly 16 years, in which
time its US-dollar price has moved from cents to several tens of
thousands of dollars --- a span of approximately six orders of
magnitude that is unusual for any traded asset. Daily prices are
publicly observable, the supply schedule is fixed by protocol, and
the network has accumulated a settlement history of
several hundred million transactions. These features make
Bitcoin a natural object of study for the kind of empirical
question that econometrics rarely gets to ask of a market
instrument: whether a single, well-defined functional law describes
its long-run growth.

\begin{figure}[!t]
\centering
\includegraphics[width=0.85\textwidth]{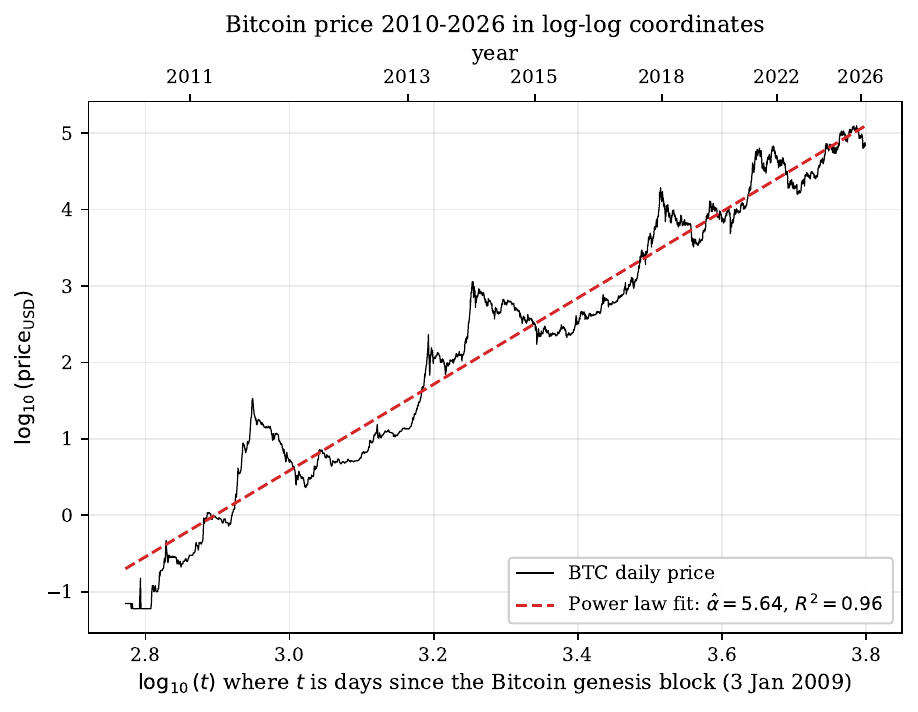}
\caption{Bitcoin daily closing price from 18 August 2010 to 25 March
2026 ($n = 5{,}699$) on log--log axes. Time is days since the
Bitcoin genesis block (3 January 2009); the upper axis shows
calendar years for reference. The dashed red line is the
power-law fit $\log_{10}(P) = \alpha\,\log_{10}(t) + c$
($\hat\alpha = 5.64$, $R^2 = 0.96$). The data lies close to the
fitted line across nearly six orders of magnitude in price.}
\label{fig:motivation}
\end{figure}

The empirical observation that motivates such a question is shown
in Figure~\ref{fig:motivation}: on log--log axes (price against time
since Bitcoin's genesis block in January 2009), the daily price
data lies remarkably close to a straight line over the entire
period 2010--2026. A straight line in log--log coordinates is the
defining signature of a power law (PL) $P \sim t^{\beta}$, with
the slope giving the exponent $\beta$. The visual fit is striking ---
$R^2 = 0.96$ on the OLS regression, with residual standard deviation
of $0.30$ in $\log_{10}(P)$ corresponding to a typical
multiplicative error of about a factor of two on price.

This empirical regularity has not gone unnoticed. A power-law
description of Bitcoin's price has been proposed in successive
forms over the last decade, with the most recent formal account in
\citet{SantostasiPerrenod2026}, who report $\hat\beta = 5.690$ over
a comparable window and propose a mechanistic derivation in which
the exponent decomposes as
$\beta = \beta_A \cdot \beta_M$, with $\beta_A$ describing
network adoption growth and $\beta_M$ a generalised Metcalfe
relationship between price and network size. A related multi-component proposal by \citet{Wheatley2019}
couples a generalised-Metcalfe value model to a Log-Periodic
Power Law Singularity bubble model. The same visual log--log
linearity is also used in popular discourse as the basis for
long-horizon valuation projections, with central estimates that
have practical consequences for portfolio allocation; the
underlying empirical claim therefore deserves the same level
of statistical scrutiny as any other quantitative forecast.

The temptation to take such projections at face value is not
specific to Bitcoin: power-law claims of this form recur across
many domains of physics, biology, and the social sciences, and
they have a well-documented history of overclaiming.
\citet{Mitzenmacher2004} surveys the algebraic relationship
between power laws and lognormal distributions and notes that
visual log--log linearity over many orders of magnitude can
arise from either family, often indistinguishably. More
strikingly, when the rigorous testing protocol of
\citet{Clauset2009} is applied to large samples of empirical
distributions previously labelled as power-law, only a small
fraction survives: \citet{BroidoClauset2019} apply the protocol
to about 1{,}000 real-world networks and find that $88\%$ are
fit at least as well by lognormal as by a power law.

Visual log--log linearity is therefore a necessary but not
sufficient condition for a structural power law. To conclude that
Bitcoin's price--time relationship \emph{is} a power law in the
sense the term has in physics --- a structural invariant of the
generating process, with a well-defined and shift-invariant
exponent --- requires more than the OLS fit in
Figure~\ref{fig:motivation}. \citet{SantostasiPerrenod2026}
explicitly acknowledge this in their Section~11.3, noting that
``the OLS estimates assume that the power law holds throughout the
entire range; likelihood-ratio tests for the lower cutoff
[\citealt{Clauset2009}] have not been applied to the temporal
series.'' This paper addresses that gap where the
Clauset--Shalizi--Newman protocol applies directly --- to Bitcoin's
tail-relevant distributional series --- and develops diagnostic
analogues for the time-domain regression where the canonical
protocol does not mechanically apply, supplementing both with a
cross-asset comparison and an out-of-sample forecasting evaluation.

\paragraph{Contributions.}
This paper makes four contributions:

\begin{enumerate}[leftmargin=2em]
\item \textbf{Distributional CSN tests.} We apply the four-step
CSN protocol to Bitcoin's tail-relevant random variables --- the
cross-section of UTXO balances at yearly snapshots, and the
marginal distribution of daily $|$returns$|$ on the full sample
and three sub-periods. Eight of eleven tests reject the power-law
hypothesis at $p < 0.05$, with seven of those at the bootstrap
resolution limit; lognormal is preferred decisively in every
case where the test has sufficient power
(Section~\ref{sec:distributional}).

\item \textbf{Time-domain CSN adaptations.} We develop three
principled adaptations of the CSN protocol for the time-domain
regression $P \sim t^\beta$: a shift-parameter sensitivity sweep
analogous to CSN's $x_{\min}$ selection; a bootstrap goodness-of-fit
test on residual diagnostics; and Vuong's likelihood-ratio test
against alternative trend specifications. The first shows that
the fitted exponent varies by nearly a factor of three across
reasonable shifts; the second shows that standard residual
diagnostics do not discriminate a power law plus realistic noise
from alternative trend models; the third, supplemented by a
parameter-matched flexibility check, shows that a 3-component
sigmoid stack wins the in-sample comparison even against
flexibility controls of equal parameter count
(Section~\ref{sec:time-domain}).

\item \textbf{Cross-series structural comparison.} We compare
Bitcoin price against eight reference series --- five Bitcoin
on-chain metrics (hash rate, addresses, transactions, difficulty,
UTXO count) and three major traditional asset classes (NASDAQ
Composite, S\&P~500, gold) --- in an in-sample single-component
comparison across all nine series (whether any of exponential,
single sigmoid, quadratic, or polynomial improves on the power
law), and a quarterly $K = 3$ wave-stability bootstrap on a
seven-series subset (Bitcoin price, hash rate, NASDAQ, S\&P~500,
gold, plus Ethereum and the Global X Lithium ETF as additional
candidate adoption-driven assets; the four on-chain metrics
other than hash rate are omitted from the bootstrap for
computational cost). \emph{Bitcoin price is the only series in
the in-sample test where no single-component specification
improves on the power law, and the only series with two stable
components at the strict 15\%-CV threshold in the
wave-stability bootstrap; it rejects the PL+AR(1) null at
$p_{<15\%} = 0.015$.} Two methodologically independent tests
therefore converge on Bitcoin price as the distinctive series
in the comparison set (Section~\ref{sec:waves}).

\item \textbf{Out-of-sample forecasting evaluation.} We evaluate
all candidate models, plus four standard time-series baselines
(RW with drift, auto-ARIMA, ETS, local-linear-trend
state-space), in walk-forward
validation across 11 yearly cutoffs (2014--2024) at horizons of
1, 3, 6, 12, 18, and 24 months, with formal pairwise
Diebold--Mariano significance. The multi-sigmoid that wins the
in-sample comparison is among the worst at long horizons; the
no-skill Naive baseline owns 1--3 month horizons; the simple
power law owns 12--24 month horizons against every parametric
trend specification \emph{and} every standard time-series
baseline at $p < 0.05$.
The fit-prediction tradeoff is the practical counterpart of the
descriptive findings (Section~\ref{sec:forecasting}).
\end{enumerate}

\paragraph{Roadmap.}
Section~\ref{sec:related} reviews the relevant literature on
power-law theory and testing, on prior Bitcoin price models, and
on out-of-sample forecast evaluation.
Section~\ref{sec:methodology} defines the data, candidate
trend specifications, evaluation metrics, and statistical tests
in detail; the methodology section is deliberately self-contained
and accessible to a reader with general mathematical training but
no specialised background in forecasting or in power-law
statistics. Sections~\ref{sec:distributional}--\ref{sec:forecasting}
report the four sets of results outlined above. Section~\ref{sec:discussion}
synthesises the findings and distinguishes structural description
from out-of-sample prediction. Section~\ref{sec:conclusion}
summarises and identifies questions left open by this work.

\section{Related work}
\label{sec:related}

This paper sits at the intersection of three literatures: rigorous
testing of empirical power-law claims; Bitcoin price models in the
time domain; and out-of-sample forecast evaluation in financial
time series. We review each in turn, deliberately
prioritising recent (2022--2026) work and methodological touchstones
over breadth.

\subsection{Power-law theory and rigorous testing}
\label{sec:related-pl-theory}

The canonical reference for empirical power-law testing is
\citet{Clauset2009}, which formalises a four-step protocol
(maximum-likelihood exponent estimation with optimal lower cutoff,
Kolmogorov--Smirnov bootstrap goodness-of-fit, and Vuong's
likelihood-ratio comparison against alternative distributions) and
demonstrates that visual log--log linearity over many decades is
not by itself sufficient evidence of a structural power law. Their
framework is the methodological backbone for the distributional
side of our analysis (Section~\ref{sec:distributional}) and the
template for our time-domain adaptations (Section~\ref{sec:time-domain}).

\citet{Mitzenmacher2004} surveys the algebraic relationship between
power-law and log-normal distributions, noting that small variations
in generative models (preferential attachment, multiplicative
processes, optimisation) produce either family and that
visual discrimination over wide ranges is genuinely difficult ---
the motivation for using formal statistical tests rather than OLS
on log--log axes.
\citet{BroidoClauset2019} apply the \citet{Clauset2009} protocol
to about 1{,}000 real-world networks and find that 88\% are fit at
least as well by lognormal as by a power law; the implied null
hypothesis (``most claimed power laws fail rigorous testing'') is
the methodological template for our analysis.

On the cost of confusing in-sample fit with structural truth more
generally, \citet{Bailey2014} formalise the relationship between
backtest overfitting and out-of-sample performance: they prove
that under reasonable assumptions, in-sample optimisation
\emph{actively degrades} expected out-of-sample accuracy when the
model-selection trial space is not disclosed. We do not engage
this literature directly, but the principle motivates our
parameter-matched flexibility check (Section~\ref{sec:in-sample-results})
and our walk-forward design (Section~\ref{sec:forecasting}).

\subsection{Bitcoin price models and the time-domain power law}
\label{sec:related-bitcoin-models}

The most recent formal account of the Bitcoin price--time
relationship is \citet{SantostasiPerrenod2026}. They report
$\hat\beta = 5.69$ over a window comparable to ours
(Section~\ref{sec:data}), develop a mechanistic derivation in
which the exponent decomposes as
$\beta = \beta_A \cdot \beta_M$ (with $\beta_A$ for network
adoption growth and $\beta_M$ a generalised Metcalfe relationship
between price and network size), and validate the empirical
regularity through four scale-invariance tests --- a pair-ratio
test, a direct collapse test, a rolling-window stability test, and
a Bayesian sequential update on local-slope estimates. Their
analysis is methodologically careful within its OLS-regression
framework. They explicitly acknowledge in their Section~11.3 that
``likelihood-ratio tests for the lower cutoff
[\citealt{Clauset2009}] have not been applied to the temporal
series.'' Our paper addresses that gap where the protocol applies
directly --- to Bitcoin's tail-relevant distributions
(Section~\ref{sec:distributional}) --- and develops diagnostic
analogues for the time-domain regression where it does not
(Section~\ref{sec:time-domain}), and reproduces all four of their
scale-invariance tests on a synthetic 3-component sigmoid stack
to characterise what those tests can and cannot demonstrate
(Section~\ref{sec:scale-invariance}).

Two related multi-component proposals frame complementary aspects
of Bitcoin's price dynamics. \citet{Wheatley2019} couple a
generalised Metcalfe value model (with logistic active-user growth
and exponent $\beta \approx 1.69$) to a Log-Periodic Power Law
Singularity bubble model, identifying a universal super-exponential
signature across four bubbles and providing ex-ante probabilistic
crash brackets. \citet{RuddPorter2025} propose a logistic
supply-and-demand framework for Bitcoin price forecasting that
explicitly substitutes a saturating diffusion process for time-only
power-law extrapolation. Both are consistent with our finding that Bitcoin's price
contains multi-component saturation dynamics, though neither
tests the alternative against the power law via the procedures
we use.

The closest published precedent for out-of-sample rejection of a
time-domain power-law specification on Bitcoin is
\citet{Shelton2024}. Using the Welch--Goyal / Rapach
walk-forward methodology with Campbell--Thompson $R^2_{OOS}$ and
Clark--West tests, the paper evaluates fifteen Bitcoin return
predictors --- including $\log(\text{time})$ --- across multiple
sub-periods. The $\log(\text{time})$ predictor has in-sample
$R^2 = 0.002$ and OOS $R^2 = -0.070$ (worse than a no-skill
mean), and the Stock-to-Flow specification of \citet{Shanaev2019}
is shown to be 80.6\% correlated with $\log(\text{time-since-Genesis})$
and to lose significance under time fixed effects. Our paper
extends this OOS rejection from the returns to the level setting
and complements it with the structural CSN tests that
\citet{Shelton2024} does not apply.

The earlier econometric critique by \citet{Shanaev2019} establishes
that Bitcoin valuation regressions on hashrate and on transactions
are spurious under instrumental-variable / endogeneity-corrected
estimation, and that the high $R^2$ values typically reported in
these specifications reflect common stochastic trends rather than
structural relationships. We use \citet{Shanaev2019} as part of the
methodological motivation for first-differencing in our causality
analysis (Section~\ref{sec:causality}).

\subsection{Cryptocurrency forecasting and regime models}
\label{sec:related-forecasting}

The forecasting literature on cryptocurrencies in general and
Bitcoin in particular is extensive; we cite a representative
sample. \citet{Catania2019} establish the methodological
framework for short-horizon crypto forecasting in
\emph{International Journal of Forecasting} using dynamic model
averaging and the Model Confidence Set procedure of
\citet{DieboldMariano1995} on 1--7 day return forecasts.
\citet{Gradojevic2023} apply random forests with technical
indicators to Bitcoin daily and hourly returns, finding
significant outperformance of the random walk only at the daily
horizon and identifying multiple structural breakpoints in
predictive importance over 2015--2019.
\citet{Gurgul2025} integrate transformer-based NLP, on-chain
metrics, and traditional financial signals in a deep learning
framework for short-horizon BTC and ETH forecasting.

For longer-horizon and regime-switching approaches more directly
related to our multi-component framing,
\citet{ForoutanLahmiri2025} apply Bayesian MCMC covariate
selection inside hidden Markov models with sixteen macroeconomic
and on-chain factors over 2016--2024; they report 30-step-ahead
MAPE values often exceeding 100\%, providing independent
methodological corroboration of our finding that long-horizon
Bitcoin point forecasts are inherently difficult.
\citet{OpreaBara2026} train a Gaussian HMM on log-returns to
infer bull/bear/sideways regimes and route each regime to a
specialised forecaster, an example of regime-switching
methodology applied to Bitcoin returns.

Two papers provide direct support for the no-skill-baseline half
of our forecasting result. \citet{MagnerHardy2022} establish a
``Meese--Rogoff puzzle'' for cryptocurrencies via Wild
Clark--West nested-model tests and rolling-window evaluation,
finding that random-walk-with-drift baselines are difficult to
beat for daily crypto returns.
\citet{Puoti2024} use Complexity-Entropy plane and Power
Spectral Density analysis to show that BTC and four other major
cryptocurrencies behave like Brownian noise (PSD $\sim 1/f^2$),
and that across fourteen statistical and ML forecasters the simple
Naive baseline ties or beats every alternative at horizons of 1,
7, and 30 days. Their finding mirrors and corroborates our
short-horizon walk-forward result (Section~\ref{sec:forecasting})
via an independent methodological route.

Finally, \citet{Tekin2024} applies Bai--Perron multiple
structural-break tests to Bitcoin and Ethereum return and variance
series, identifying common breakpoints aligned with COVID-era and
post-pandemic shocks. The use of formal break-point econometrics
on cryptocurrency series provides an empirical precedent for the
structural-break methodology that our wave-stability bootstrap
generalises.

\subsection{Out-of-sample evaluation methodology}
\label{sec:related-oos}

The methodological framework for our walk-forward forecasting
evaluation rests on three foundational references.
\citet{MeeseRogoff1983} establish the canonical
``naive-beats-structural'' result for exchange rates, showing that
the random walk dominates estimated structural macro-finance
models at 1--12 month horizons even when the structural models
are given the unfair advantage of realised future explanatory
variables. The result is one of the most replicated negative
findings in empirical macroeconomics and is the natural
intellectual precedent for our short-horizon Naive result on
Bitcoin (Section~\ref{sec:forecasting}).
\citet{Tashman2000} surveys out-of-sample evaluation practice in
forecasting and codifies the walk-forward design (sequential
training-set extension with origin-fixed forecast horizons) we
adopt.
\citet{DieboldMariano1995} introduce the formal pairwise test for
equal predictive accuracy that we use throughout
Section~\ref{sec:forecasting}; we apply the
heteroscedasticity- and autocorrelation-consistent variance
correction of \citet{NeweyWest1987} for the residual
autocorrelation induced by overlapping training windows.

The statistical hazards of regressing one non-stationary series on
another are formalised in \citet{GrangerNewbold1974}, whose
spurious-regression result motivates our use of first-differencing
in the causality analysis (Section~\ref{sec:causality}). The
augmented Dickey--Fuller stationarity tests we cite there derive
from \citet{DickeyFuller1979} and \citet{SaidDickey1984}.

\section{Data and methodology}
\label{sec:methodology}

This section defines the data sources, candidate trend specifications,
evaluation metrics, and statistical tests used throughout the paper.
The presentation is deliberately self-contained: each metric and test
is introduced with intuition, formula, and interpretation guide before
any empirical results appear. A specialist reader can skim; a reader
with general mathematical training but no specific background in
forecasting or in power-law statistics has, by the end of this
section, the toolkit needed to read the rest of the paper.

\paragraph{Notation: $\alpha$ versus $\beta$.}
We use $\alpha$ for the exponent fitted in our OLS time-domain
regression $\log_{10}(P) = \alpha\,\log_{10}(t) + c$, following
standard regression-coefficient convention. Earlier work
\citep{SantostasiPerrenod2026, Wheatley2019} writes the same
quantity as $\beta$, motivated by a generative reading of
$P \sim t^{\beta}$. We retain $\beta$ when paraphrasing those
references, and the two symbols denote the same scalar wherever
they appear; the numerical values $\hat\alpha = 5.644$ (this paper)
and $\hat\beta = 5.690$ \citep{SantostasiPerrenod2026} differ only
because of slightly different windows
(Section~\ref{sec:data}). One unrelated occurrence of $\beta$ is
the curvature parameter inside the stretched-exponential family
$\log_{10}(P) = a + b\,t^{\beta}$
(Section~\ref{sec:candidates}); the context disambiguates.

\subsection{Data}
\label{sec:data}

The primary series is the daily closing price of Bitcoin in US dollars
from 18 August 2010 to 25 March 2026, $n = 5{,}699$ observations,
obtained from the CoinDesk Bitcoin Price Index. We exclude the period
before mid-2010 because exchange data is sparse and unreliable; the
first publicly traded prices on liquid exchanges appear in this
period. Time is measured in days since Bitcoin's genesis block,
$t_0 = 3$~January 2009. Throughout the paper, we use $t$ to denote
elapsed days since $t_0$, so the first observation has $t \approx 592$
and the last $t \approx 6{,}291$.

We complement the price series with five on-chain metrics, daily,
covering 18 August 2010 to 16 March 2026 ($n \approx 5{,}690$): hash
rate (estimated network mining capacity), the number of unique
addresses with non-zero balance, the daily count of confirmed
transactions, mining difficulty, and the unspent transaction output
(UTXO) count. These metrics are computed from a full archival Bitcoin
node operated by the authors, with the entire blockchain processed
locally; no third-party aggregator (Blockchain.com, Glassnode,
Coin~Metrics, etc.) is involved. This avoids any risk of
inconsistencies between aggregators in how addresses are deduplicated,
how UTXOs are counted at chain reorganisations, or how hash rate is
estimated from block-time variance. The on-chain window aligns
closely with the price window; minor end-date differences (one to
two days) reflect block-time variation and have no bearing on the
analyses below.

For cross-asset comparison we use daily closing prices of three
benchmark assets covering the same calendar window as Bitcoin price
(2010-08-18 to 2026-03-25): the NASDAQ Composite, the S\&P~500, and
gold (continuous front-month futures). Trading-day closures yield
$n \approx 3{,}924$ for each benchmark --- noticeably smaller than
Bitcoin's $n = 5{,}699$ because traditional exchanges close on
weekends and public holidays, whereas Bitcoin trades continuously.
Cross-asset prices are taken from Yahoo Finance.

\paragraph{Time from the genesis block.} The genesis block is a
natural absolute reference for Bitcoin: it predates any market data,
is unambiguous, and matches the convention used by every prior paper
that proposes a power-law relationship $P \sim t^\beta$. For
cross-asset comparison we apply the same time origin: the relative
positions of NASDAQ, S\&P, and gold within the window are
identical to Bitcoin's, so any difference in the fitted exponents
reflects the data, not the time coordinate.

\paragraph{Comparison with prior data windows.} Our window
(2010-08-18 to 2026-03-25, $n = 5{,}699$) is closely comparable to
that used by \citet{SantostasiPerrenod2026} (2010-07-18 to
2026-02-18, $n = 5{,}696$, with an explicit exclusion of
$t < 400$~days that is automatically satisfied by the data). Both
windows omit pre-mid-2010 data for liquidity reasons. The slight
difference at each endpoint accounts for the $\sim 0.8\%$ gap
between our fitted exponent ($\hat\alpha = 5.644$) and theirs
($\hat\beta = 5.690$); we discuss this further in
Section~\ref{sec:time-domain}.

\subsection{Candidate trend specifications}
\label{sec:candidates}

This subsection lists the candidate forms used for the
\emph{time-domain regression} $\log_{10}(P) = f(t) + \varepsilon$,
i.e.\ all candidate trend models in $t$.
These are the alternatives compared in
Sections~\ref{sec:time-domain} and \ref{sec:waves}.
A separate set of candidate \emph{distributions}
(lognormal, exponential-as-a-distribution, stretched-exponential-as-a-distribution,
power law with exponential cutoff) is used in the distributional
tests of Section~\ref{sec:distributional}; those alternatives are
introduced as part of the CSN protocol in
Section~\ref{sec:csn-protocol} (step~4).

The candidate set is organised by motivation:

\begin{itemize}[leftmargin=2em]
\item \emph{No-skill baseline.} The naive forecast establishes the
floor of forecasting skill: any model claiming predictive value must
beat it, and at short horizons it is famously hard to beat for
asset prices \citep{MeeseRogoff1983}.
\item \emph{Power law and closely related smooth specifications.}
The power law has been proposed for Bitcoin's price--time relationship
in successive forms; we adopt the most recent formalisation by
\citet{SantostasiPerrenod2026}, who report
$\hat\beta \approx 5.69$ over a comparable window. The pure
exponential is included for completeness, since it is the strict
form of the casually invoked claim that ``Bitcoin grows
exponentially'' (a claim our analysis will show holds for
NASDAQ and S\&P 500 over our window but not for Bitcoin). The
stretched exponential is informative because its $\beta \to 0$
limit algebraically reduces to a power law, providing a natural
test of smooth-model equivalence. Quadratic and cubic in $t$ are
included as generic polynomial competitors.
\item \emph{Saturation-curve alternatives.} The logistic (sigmoid)
curve is the standard model for diffusion-of-innovations adoption
phenomena \citep{Bass1969}; a stack of $K$ sigmoids generalises this
to multiple successive adoption events. Multi-component variants
of this idea appear in earlier work on Bitcoin price dynamics in
different forms --- for example, \citet{Wheatley2019} couple a
generalised-Metcalfe value model (with logistic active-user growth)
to a log-periodic-power-law-singularity bubble model. Our $K=1$
case is the simplest single-saturation specification; $K=3$ is the
smallest stack that can capture multi-wave behaviour while remaining
parsimonious.
\item \emph{Parameter-matched flexibility controls.} Polynomial of
degree~9 (with $t$ rescaled to $[-1, 1]$ for numerical stability)
and a cubic B-spline with ten basis functions are not motivated by
any specific theory; they have the same parameter count as the
$K=3$ multi-sigmoid and serve as flexibility controls. If $K=3$
wins in-sample only because ten parameters can fit any reasonable
curve, both controls should match its performance; if $K=3$ wins
because the sigmoid functional form captures structural
information, the controls should fall short.
\item \emph{Standard forecasting baselines.} For
out-of-sample evaluation we additionally include four baselines that
are conventional in the time-series forecasting literature: random
walk with drift, automatically selected ARIMA, exponential smoothing
with additive trend (ETS), and a local-linear-trend
unobserved-components state-space model. Unlike the trend
specifications above, these baselines do not posit a deterministic
parametric form for $\log_{10}(P)$ as a function of $t$; they
extrapolate from the recent dynamics of the level series. They are
included to assess whether the power law's long-horizon advantage
in Section~\ref{sec:forecasting} is robust to standard
time-series alternatives.
\end{itemize}

The detailed forms used throughout the paper are:

\begin{description}[leftmargin=2em, style=nextline]
\item[Naive (no skill):]
$\hat P(t+h) = P(t)$. The forecast at any horizon equals the most
recent observed price. Two parameters' worth of effective
information (the last value); used as the no-skill baseline for
out-of-sample evaluation.

\item[Power law (PL):]
$\log_{10}(P) = \alpha \, \log_{10}(t) + c$. A straight line in
log--log coordinates; the exponent $\alpha$ measures the percent
change in $P$ per percent change in $t$. Two parameters.

\item[Pure exponential:]
$\log_{10}(P) = a\,t + b$. A straight line in semi-log coordinates;
$P$ doubles every fixed time interval $\ln 2 / a$. Two parameters.

\item[Quadratic and cubic in $t$:]
$\log_{10}(P) = c + a\,t + b\,t^2$ (three parameters) and the
extension with a $t^3$ term (four parameters). Polynomial
generalisations of the exponential.

\item[Stretched exponential:]
$\log_{10}(P) = a + b\,t^{\beta}$. A flexible curve that
interpolates between exponential ($\beta = 1$), linear-in-$t$
($\beta \to 0$ in the appropriate limit), and faster-than-exponential
growth. Three parameters.

\item[Sigmoid (logistic):]
$\log_{10}(P) = b + L \,/\, [1 + e^{-k(t-t_0)}]$. An S-shaped
curve that grows from $b$ to $b + L$ with maximum slope at $t = t_0$.
Useful for adoption / saturation processes. Four parameters.

\item[Multi-sigmoid stack ($K$ components):]
$\log_{10}(P) = b + \sum_{i=1}^{K} L_i \,/\, [1 + e^{-k_i(t-t_{0,i})}]$.
A sum of $K$ logistic curves that can capture multiple successive
saturation events. We will primarily use $K=1$ and $K=3$. The model
has $1 + 3K$ parameters; $K=3$ has 10.

\item[Polynomial (degree $d$):]
$\log_{10}(P) = \sum_{j=0}^{d} c_j \tilde t^{\,j}$, where
$\tilde t \in [-1, 1]$ is $t$ rescaled to the training window. We
take $d=9$ (10 parameters) as a parameter-matched alternative to
the $K=3$ multi-sigmoid; the rescaling is needed for numerical
stability at high degree.

\item[Cubic B-spline (10 basis):]
A piecewise cubic with knots placed at quantiles of $t$ in the
training window, expressed in the cubic B-spline basis. With ten
basis functions this is also parameter-matched to $K=3$, and
provides a flexible \emph{locally} adapting alternative.

\item[Random walk with drift (RW-drift):]
ARIMA$(0, 1, 0)$ with constant: $\log_{10} P_t = \log_{10} P_{t-1}
+ \mu + u_t$ with $u_t \sim \mathcal{N}(0, \sigma^2)$. The
$h$-step forecast is $\log_{10} P_t + h\mu$. Two parameters
($\mu, \sigma^2$); the canonical baseline of asset-price
forecasting \citep{MeeseRogoff1983, MagnerHardy2022}.
\item[Auto-ARIMA:] ARIMA$(p, d, q)$ with $(p, d, q)$ selected by
AIC over $p, q \in [0, 3]$, $d \in [0, 2]$
\citep{Tashman2000} via the stepwise procedure of
\citet{HyndmanKhandakar2008} as implemented in \texttt{pmdarima}.
\item[Exponential smoothing (ETS):] Holt's linear trend
$\hat L_t = \alpha y_t + (1-\alpha)(\hat L_{t-1} + \hat T_{t-1})$,
$\hat T_t = \beta(\hat L_t - \hat L_{t-1}) + (1-\beta)\hat T_{t-1}$,
no seasonality. Three parameters ($\alpha, \beta,$ initial level/
trend); a workhorse of practical forecasting.
\item[Local linear trend (LLT):]
Unobserved-components state-space model
\citep{HarveyDurbin1986} with random level and
random slope: $y_t = \mu_t + \varepsilon_t$,
$\mu_t = \mu_{t-1} + \nu_{t-1} + \xi_t$,
$\nu_t = \nu_{t-1} + \zeta_t$, with mutually independent Gaussian
shocks. Same forecast at long horizons as a stochastic-trend extension
of ETS but with explicit variance components.

\end{description}

\subsection{In-sample evaluation metrics}
\label{sec:in-sample-metrics}

We use five complementary measures to compare candidate trend
specifications on the same data.

\paragraph{Coefficient of determination ($R^2$).}
The fraction of variance in $\log_{10}(P)$ explained by the model:
$R^2 = 1 - \mathrm{SSE} / \mathrm{SST}$, where SSE is the sum of
squared residuals and SST the total variance. Bounded between 0 and 1
and intuitive, but $R^2$ never decreases when parameters are added,
so it cannot be used alone to compare models of different
complexity.

\paragraph{Residual standard deviation ($\sigma_{\mathrm{resid}}$).}
The typical fit error in the same units as the data:
$\sigma_{\mathrm{resid}} = \sqrt{\mathrm{SSE} / n}$. Because we work
in $\log_{10}(P)$, $\sigma_{\mathrm{resid}} = 0.30$ corresponds to a
typical multiplicative error of $10^{0.30} \approx 2$ on price.

\paragraph{Log-likelihood ($\ell$).}
Under the standard assumption of Gaussian residuals, the
log-likelihood of the data given the model at the maximum likelihood
estimate is
\[
\ell = -\tfrac{n}{2} \bigl[\,\ln(2\pi) + \ln(\hat\sigma^2) + 1\bigr],
\quad \hat\sigma^2 = \mathrm{SSE} / n.
\]
Higher is better. Like $R^2$, $\ell$ increases monotonically with
flexibility; we use it as a building block, not as a model-selection
criterion.

\paragraph{Akaike Information Criterion (AIC).}
$\mathrm{AIC} = -2\ell + 2k$, where $k$ is the number of parameters.
Lower is better. AIC estimates the expected out-of-sample prediction
error of a model under the K\"ullback--Leibler divergence to the true
distribution \citep{Akaike1974}. As a working interpretation,
differences in AIC are read on the following scale:

\begin{center}
\begin{tabular}{rl}
\toprule
$\Delta\mathrm{AIC}$ & interpretation \\
\midrule
$<2$ & models essentially equivalent \\
$2$--$7$ & clear preference for the lower-AIC model \\
$>10$ & decisive preference \\
\bottomrule
\end{tabular}
\end{center}

\paragraph{Bayesian Information Criterion (BIC).}
$\mathrm{BIC} = -2\ell + k\,\ln n$. Same form as AIC but with a
heavier parameter penalty: for our $n$, the BIC penalty per parameter
is $\ln n \approx 8.6$, more than four times the AIC penalty of $2$.
Lower is better. AIC is asymptotically efficient (selects the model
with best prediction); BIC is consistent (selects the true model if
it is in the candidate set). When the two criteria agree, the
conclusion is robust to the choice of complexity penalty. We report
both throughout.

\paragraph{Reporting multiple criteria.}
$R^2$ and $\sigma_{\mathrm{resid}}$ are intuitive descriptors but
biased toward flexible models. AIC and BIC penalise complexity in
different ways, and their agreement is a sensitivity check on any
model selection. The log-likelihood is the underlying statistical
quantity from which AIC, BIC, and the formal tests below are
derived.

\subsection{Vuong's likelihood-ratio test}
\label{sec:vuong}

For two non-nested models $A$ and $B$ fit to the same data, Vuong's
test \citep{Vuong1989} formally tests whether one provides a
significantly better fit. Let $\ell_A^{(i)}$ and $\ell_B^{(i)}$
denote the pointwise log-likelihood contributions for observation
$i$. The test statistic is
\begin{equation}
\label{eq:vuong}
V \;=\; \frac{\sqrt{n}\,\bar{\mathrm{LR}}}{s_{\mathrm{LR}}},
\qquad
\mathrm{LR}_i = \ell_A^{(i)} - \ell_B^{(i)},
\end{equation}
where $\bar{\mathrm{LR}}$ and $s_{\mathrm{LR}}$ are the sample mean
and standard deviation of $\mathrm{LR}_i$. Under the null hypothesis
that the two models are equivalent, $V$ is asymptotically standard
normal, so $|V| > 1.96$ corresponds to $p < 0.05$. A positive $V$
means model $A$ wins; negative means $B$ wins.

When the two candidates have different numbers of parameters, we use
the BIC-corrected variant
$\bar{\mathrm{LR}}_{\mathrm{corrected}} = \bar{\mathrm{LR}} +
(k_A - k_B)\,\ln n / (2n)$, which penalises the more flexible model.
This adjustment is conservative: it can only attenuate the apparent
advantage of the flexible model.

\paragraph{Where Vuong's test is used in this paper.} The test
predates the CSN protocol (Section~\ref{sec:csn-protocol}) by two
decades and is a general-purpose tool, but CSN adopted it as their
fourth step. We use it in both senses, and introduce it once here
to avoid re-defining it. First, as Step~4 of the canonical CSN
protocol on the distributional tests of Section~\ref{sec:distributional}
(power law against lognormal, exponential, and stretched exponential
on UTXO balances and daily~$|$returns$|$). Second, in our
time-domain CSN adaptation C (Section~\ref{sec:csn-adaptations}),
applied directly to the time-domain regression to compare the power
law against alternative trend specifications such as the
multi-sigmoid stack.

\subsection{The Clauset--Shalizi--Newman (CSN) protocol}
\label{sec:csn-protocol}

Power-law claims are notoriously hard to verify by visual inspection.
Log-normal, stretched-exponential, and other heavy-tailed
distributions can produce convincing log--log linearity over many
decades \citep{Mitzenmacher2004}; many published power-law claims
across disciplines fail formal testing once the claim is examined
rigorously \citep{BroidoClauset2019}. \citet{Clauset2009} developed
a four-step protocol for testing such claims that has become
canonical:

\begin{enumerate}[leftmargin=2em]
\item \textbf{Estimate the lower cutoff $x_{\min}$} above which the
power law is claimed to hold, by minimising the
Kolmogorov--Smirnov (KS) distance between the fitted power law and
the empirical CDF for $x \geq x_{\min}$.
\item \textbf{Estimate the exponent $\hat\alpha$ by maximum
likelihood}, not by log--log regression. The MLE is more efficient
and free of the bias of binned regression.
\item \textbf{Compute a bootstrap goodness-of-fit p-value.} Generate
many synthetic datasets of size $n$ from the fitted power law with
the estimated $x_{\min}$; fit a power law to each (with its own
$x_{\min}$ selection) and record the KS distance. The p-value is
the fraction of synthetic KS distances $\geq$ the observed KS
distance. Values $> 0.10$ indicate the power law is plausible;
values $< 0.10$ reject the power law.
\item \textbf{Compare against alternative distributions} (lognormal,
exponential, stretched exponential, power law with exponential
cutoff) using Vuong's likelihood-ratio test.
\end{enumerate}

The CSN protocol applies natively to \emph{distributional} power
laws --- claims that some random variable $X$ has
$P(X \geq x) \sim x^{-\alpha}$. We apply it directly to two such
cases for Bitcoin: the distribution of UTXO balances (yearly
snapshots) and the distribution of daily $|$returns$|$
(Section~\ref{sec:distributional}).

The time-domain claim that $P \sim t^\beta$ is a regression
statement, not a distributional one. Time $t$ is not an IID sample
from some distribution but an ordered index, and the relationship
asserts a deterministic trend with noise rather than the tail
behaviour of a random variable. The CSN procedure cannot be applied
mechanically in this setting; we develop principled adaptations
next.

\subsection{Time-domain CSN adaptations}
\label{sec:csn-adaptations}

We adapt three of the four CSN steps to the time-domain regression
$\log_{10}(P) = \alpha\,\log_{10}(t + s) + c$, where $s \geq 0$ is
a shift parameter (described next). Step~2 of the canonical
protocol --- maximum-likelihood estimation of the exponent --- is
not separately adapted because under Gaussian residuals OLS on the
time-domain regression is itself the maximum-likelihood estimator.
The remaining three steps (lower-cutoff selection, bootstrap
goodness-of-fit, and likelihood-ratio comparison against
alternatives) are adapted as Adaptations~A, B, C below.

\paragraph{Diagnostic analogues, not formal CSN extensions.}
The canonical CSN protocol is derived for IID samples from a
distribution. The time-domain regression is a different statistical
object: an ordered, deterministic-trend regression with autocorrelated
residuals. We therefore frame Adaptations~A--C as \emph{diagnostic
analogues} rather than formal extensions: each takes a CSN-style
question (does the fit depend on a free parameter? are the
residuals more extreme than the null implies? does an alternative
specification beat it on likelihood?) and adapts it to the
time-domain setting. The diagnostics inherit the CSN protocol's
spirit --- that visual fit is not enough --- but their statistical
guarantees are weaker than CSN's distributional results.
Specifically, the bootstrap goodness-of-fit (Adaptation~B) and the
wave-stability bootstrap (Section~\ref{sec:wave-stability}) require
a model of the residual process; we use AR(1) calibrated to the
data, with alternative null choices (block bootstrap, ARMA/GARCH
residual models) briefly considered in
Section~\ref{sec:bootstrap}.

The shift parameter $s$ controls where the power-law time origin
sits \emph{relative to the genesis block}: $s = 0$ places the
power-law origin at the genesis block itself, while $s > 0$ shifts
the implied origin $s$~days into the past. The shift is a
methodological choice independent of when the observed data
begins: our window starts at $t \approx 592$~days because Bitcoin
was not liquidly traded before mid-2010, but the power-law
$t$-coordinate is anchored to the genesis block and runs from
$t = 0$ conceptually, with $s$ controlling any offset of the fit's
reference. The default specification used throughout this paper
is $s = 0$, matching \citet{SantostasiPerrenod2026}; its
empirical justification is given in
Section~\ref{sec:time-domain}.

\paragraph{Adaptation A: shift-parameter sensitivity (xmin analog).}
Sweep $s$ over a grid (here, $s \in [0, 5000]$ days) and record
$\hat\alpha(s)$, $R^2(s)$, $\mathrm{AIC}(s)$. An exponent that is
robust to specification (a necessary condition for a structural
reading) should be approximately invariant to $s$ over a reasonable
range; large variation indicates that the apparent law depends on a
methodological choice rather than on a property of the data alone.

\paragraph{Adaptation B: bootstrap goodness-of-fit on regression
diagnostics.} Generate $B$ synthetic trajectories under the null
that the data is generated by the fitted power law plus
autocorrelated noise calibrated to the observed residuals (the
noise model is described in Section~\ref{sec:bootstrap}). For each
synthetic trajectory, compute the same residual diagnostics as on
the real data (e.g., the Ljung--Box autocorrelation test
\citep{LjungBox1978}, the Breusch--Pagan heteroscedasticity test
\citep{BreuschPagan1979}, Chow break-test counts \citep{Chow1960}).
The p-value is
the fraction of synthetic trajectories that produce diagnostics as
extreme as the observed. If the observed diagnostics fall within
the synthetic distribution, the standard pathologies are
non-discriminating: they are reproduced under the very null they
would be used to reject.

\paragraph{Adaptation C: Vuong's likelihood-ratio test against
alternative trend specifications.} Compare the power law against
each candidate trend specification from Section~\ref{sec:candidates}
using Vuong's test (\ref{eq:vuong}), with BIC correction where
parameter counts differ. This is the time-domain analog of CSN
step~4.

\subsection{Bootstrap procedures}
\label{sec:bootstrap}

\paragraph{AR(1) noise model for time-domain bootstraps.}
Bitcoin's residuals from the power-law fit are highly autocorrelated:
neighbouring days' residuals are nearly identical. We model this as
a first-order autoregressive (AR(1)) process,
\[
\varepsilon_t \;=\; \rho\,\varepsilon_{t-1} + \eta_t,
\qquad \eta_t \sim \mathcal{N}(0, \sigma_\eta^2),
\]
with $\rho$ and the marginal residual variance
$\sigma^2 = \sigma_\eta^2 / (1 - \rho^2)$ estimated from the
observed PL residuals. For Bitcoin price at $s=0$ the calibration
gives $\hat\rho \approx 0.998$ and $\hat\sigma \approx 0.302$
(reported in Section~\ref{sec:time-domain}).

\paragraph{Choice of null.}
AR(1) is one parametric null among several plausible models for
log-price residuals; alternatives include block bootstraps and
ARMA/GARCH residual models that capture volatility clustering.
With $\hat\rho = 0.998$ the AR(1) is close to the unit-root limit
and inherits much of the heavy-tail and clustering behaviour of
the empirical residuals through its near-random-walk persistence.
We retain AR(1) as the primary specification because it is the
simplest parametric form that matches the dominant autocorrelation
feature of the data.

\paragraph{Bootstrap p-value.}
For any test statistic $T$ computed on real data, we generate $B$
synthetic trajectories under a specified null hypothesis (typically:
PL plus AR(1) noise), compute $T$ for each, and report the bootstrap
p-value
\[
\hat p \;=\; \frac{\#\{\,T_b^{\,*} \geq T_{\mathrm{obs}}\,\}}{B}.
\]
Small $\hat p$ ($\leq 0.05$ at the conventional level) means the
observed statistic is unusual under the null, and we reject the null.
Bootstrap p-values are reported with ``$p < 1/B$'' wherever the
observed statistic exceeds every synthetic value.

\subsection{Out-of-sample evaluation}
\label{sec:oos}

In-sample fit quality is a measure of description, not prediction.
We evaluate predictive accuracy using \emph{walk-forward validation}
\citep{Tashman2000}: at each of 11 yearly cutoff dates from
1~January~2014 through 1~January~2024, we fit each model on data
strictly before the cutoff and forecast the price at horizons of 1,
3, 6, 12, 18, and 24 months ahead. This produces 11 forecast errors
per (model, horizon) pair on $\log_{10}(P)$, with no information
leakage from future to past.

\paragraph{Naive baseline.}
The naive forecast is $\hat P(t+h) = P(t)$ --- next month's price
predicted as today's. Because Bitcoin's daily returns have low
predictability, naive is surprisingly competitive at short horizons,
as established for traditional asset classes by
\citet{MeeseRogoff1983}. Including it as a baseline gives a
calibration of what ``forecast skill'' actually looks like.

\paragraph{Diebold--Mariano test.}
For two models $A$ and $B$ with squared forecast-error series
$e_A^2, e_B^2$ across cutoffs, the Diebold--Mariano (DM) test
\citep{DieboldMariano1995} formally tests equal predictive accuracy.
Let $d_t = e_{A,t}^2 - e_{B,t}^2$; then
\begin{equation}
\label{eq:dm}
\mathrm{DM} \;=\; \frac{\bar d}{\sqrt{\widehat{\mathrm{Var}}(\bar d)}},
\end{equation}
where the variance estimate uses the heteroscedasticity- and
autocorrelation-consistent (HAC) correction of
\citet{NeweyWest1987}. This correction is needed because forecast
errors at adjacent cutoffs may be correlated --- in our setting,
successive training windows overlap, so a 1-year cutoff increment
shares 90\%+ of training data with its neighbour; without the
correction the variance of $\bar d$ would be underestimated and
DM rejection rates inflated. We use a HAC lag of at most~2 given
our 11 cutoffs. Under the null of equal accuracy, DM is
asymptotically standard normal; we report two-sided p-values.
With $n = 11$ cutoffs per horizon, statistical power is moderate;
we therefore report effect sizes (mean RMSE differences) alongside
p-values.

\subsection{Causality tests}
\label{sec:causality-methods}

To assess the direction of information flow between Bitcoin price
and each on-chain metric we use two complementary tests, both on
first-differenced log series (which are stationary by the
augmented Dickey--Fuller test at $p < 0.001$ in every case).
First-differencing converts growth rates into stationary changes,
removing the common stochastic trend that would render levels-based
regression spurious \citep{GrangerNewbold1974, Shanaev2019}.

\paragraph{Granger causality.}
\citet{Granger1969} asks whether past values of $X$ help predict
$Y$ beyond the predictive power of $Y$'s own past. We test both
directions (price $\to$ metric, and metric $\to$ price) at lag orders
of 7, 14, 30, and 60 days, reporting the F-statistic and its
p-value for each direction. The asymmetry ratio
$F_{P \to M} \,/\, F_{M \to P}$ summarises the relative strength of
the two directions: a value $\gg 1$ indicates that price changes
predict metric changes more strongly than the reverse.

\paragraph{Cross-correlation function (CCF).}
The CCF measures the linear correlation between two series at
varying time lags:
\[
\rho_{XY}(k) \;=\; \frac{\mathrm{Cov}(X_t, Y_{t-k})}{\sigma_X \sigma_Y}.
\]
A peak at lag $k > 0$ in the CCF of (price changes, metric changes)
indicates that price changes lead metric changes by $k$ days. Where
Granger reports the existence of a directional relationship, the
CCF reports its time scale.

\section{Distributional power-law tests}
\label{sec:distributional}

The CSN protocol (Section~\ref{sec:csn-protocol}) was developed to
test power-law claims about \emph{random variables}, and Bitcoin
offers two natural candidates for such tests: the cross-sectional
distribution of unspent-transaction-output (UTXO) balances, and the
marginal distribution of daily $|$returns$|$. Both are
tail-relevant: in each case a power-law tail would imply a
heavy-tailed phenomenon of the sort the canonical protocol was
built to assess. This section applies the four CSN steps to both
series, and provides a methodological baseline before we turn to
the more difficult time-domain claim in
Section~\ref{sec:time-domain}.

\subsection{Series tested}
\label{sec:dist-series}

\paragraph{UTXO balance distributions.}
For each of seven yearly snapshots from 2013 through 2025, we
extract the cross-sectional distribution of UTXO balances expressed
in BTC. The number of UTXOs grows from $\sim 8 \times 10^6$ in 2013
to $\sim 3.5 \times 10^8$ in 2025, all derived from the local
archival node (Section~\ref{sec:data}). Balances are binned into
nine logarithmic buckets from $<10^{-3}$~BTC to $>10^{3}$~BTC and
the CSN procedure applied to each yearly distribution.

\paragraph{Daily $|$return$|$ distributions.}
Daily log-returns are $r_t = \log(P_t / P_{t-1})$. We test the
distribution of $|r_t|$ on the full sample
(2010-2026, $n = 5{,}564$ after first-differencing) and on three
non-overlapping sub-periods of approximately equal length:
2010-2015 ($n = 1{,}828$), 2016-2020 ($n = 1{,}826$), and
2021-2026 ($n = 1{,}908$). The sub-period split lets us check
whether any apparent non-power-law character of the full sample is
an artefact of regime mixing: if Bitcoin's daily $|$return$|$
distribution is truly a power law within each regime but the
exponent shifts across regimes, then the full sample would fail
even though sub-periods would each pass.

\subsection{Procedure}
\label{sec:dist-procedure}

For each distribution we apply the four CSN steps from
Section~\ref{sec:csn-protocol}: (1)~lower cutoff $\hat x_{\min}$
selected by minimising the KS distance; (2)~MLE exponent
$\hat\alpha$; (3)~bootstrap goodness-of-fit $p$ using $B = 200$
synthetic samples; (4)~Vuong's likelihood-ratio test
(Section~\ref{sec:vuong}) against lognormal, exponential, and
stretched-exponential alternatives.

\paragraph{Sample-size cap.}
CSN's bootstrap step requires re-fitting the power law to each
synthetic sample. For the largest UTXO snapshot
($n \sim 3.5 \times 10^8$) running the bootstrap on the full
sample is computationally intractable. We cap the bootstrap sample
at $n = 5{,}000$, which is justified by two properties of the data:
the binned UTXO support has only nine unique values, so re-sampling
beyond that point adds no statistical information; and $n = 5{,}000$
is well above CSN's conventional minimum of $n_{\text{tail}} > 50$
above the fitted cutoff.

\subsection{Results}
\label{sec:dist-results}

Table~\ref{tab:distributional} reports the eleven tests.

\begin{table}[htbp]
\centering
\caption{CSN distributional tests on Bitcoin's tail-relevant series.
$\hat x_{\min}$ is the fitted lower cutoff; $\hat\alpha$ is the MLE
exponent; $D$ is the observed Kolmogorov--Smirnov distance; $p$ is
the bootstrap goodness-of-fit p-value with $B = 200$ iterations
(``$<0.005$'' indicates the resolution limit of $B = 200$).
$n_{\text{tail}}$ (in footnote) is the number of bootstrap-sample
points above $\hat x_{\min}$ when relevant.}
\label{tab:distributional}
\small
\begin{tabular}{lrrrrcl}
\toprule
Series & $n$ & $\hat x_{\min}$ & $\hat\alpha$ & $D$ & $p$ & Verdict \\
\midrule
\multicolumn{7}{l}{\textit{UTXO balance distributions (yearly snapshots)}} \\
2013 & $8.1\times10^{6}$ & 0.0005 & 1.56 & 0.183 & $<0.005$ & PL rejected \\
2015 & $3.4\times10^{7}$ & 0.0005 & 1.97 & 0.141 & $<0.005$ & PL rejected \\
2017 & $6.6\times10^{7}$ & 0.0005 & 1.87 & 0.144 & $<0.005$ & PL rejected \\
2019 & $1.0\times10^{8}$ & 0.0005 & 2.21 & 0.143 & $<0.005$ & PL rejected \\
2021 & $1.3\times10^{8}$ & 0.0005 & 2.33 & 0.149 & $<0.005$ & PL rejected \\
2023 & $2.1\times10^{8}$ & 0.0005 & 2.56 & 0.148 & $<0.005$ & PL rejected \\
2025 & $3.5\times10^{8}$ & 0.316  & 2.89 & 0.190 & 0.16    & Plausible$^\dagger$ \\
\midrule
\multicolumn{7}{l}{\textit{Daily $|r_t|$ distributions}} \\
Full sample 2010--2026 & 5{,}564 & 0.0149 & 2.68 & 0.060 & $<0.005$ & PL rejected \\
2010--2015              & 1{,}828 & 0.0325 & 2.96 & 0.059 & 0.17    & Plausible$^\dagger$ \\
2016--2020              & 1{,}826 & 0.0171 & 3.00 & 0.086 & 0.035   & PL rejected \\
2021--2026              & 1{,}908 & 0.0083 & 2.55 & 0.093 & $<0.005$ & PL rejected \\
\bottomrule
\end{tabular}

\medskip
\raggedright\footnotesize
$^\dagger$ Power-of-test artefact: tail truncation leaves the procedure
underpowered. UTXO 2025: $n_{\text{tail}} = 74$ above the shifted
$\hat x_{\min} = 0.316$. Daily 2010--2015: $n_{\text{tail}} = 234$.
In both cases Vuong's likelihood-ratio test (Section~\ref{sec:dist-vuong})
prefers lognormal, indicating that the failure to reject reflects
the test's power, not evidence in favour of a power law.
\end{table}

\paragraph{Headline.}
Eight of the eleven tests reject the power-law hypothesis at
$p < 0.05$, and seven of those at the $B = 200$ resolution limit of
$p < 0.005$. The two ``plausible'' results (UTXO 2025 and daily
2010-2015) are explained by power-of-test artefacts.

\paragraph{The two ``plausible'' results.}
\begin{itemize}[leftmargin=2em]
\item For UTXO 2025, $\hat x_{\min}$ shifted from $0.0005$~BTC
(the lowest bucket, used in 2013--2023) to $0.316$~BTC (the second
bucket from the top), leaving only 74 effective tail observations.
With so few unique values in the fitted region, the KS distance has
limited power to reject. The procedure cannot find evidence against
a power law when there is too little tail. This is a known
limitation \citep[§3.4]{Clauset2009} and is reported alongside
$n_{\text{tail}}$ for transparency.
\item For daily 2010--2015, the early Bitcoin period has very
heavy-tailed returns and a small effective tail
($n_{\text{tail}} = 234$). The same procedure on 2016--2020
(similar $n_{\text{tail}} = 407$) does reject at $p = 0.035$.
\end{itemize}
Neither is evidence \emph{for} a power law: with sufficient tail
samples we always reject; the two failures-to-reject reflect
truncation, not goodness of fit.

\paragraph{Sub-period regime mixing is not the culprit.}
The hypothesis that the full-sample $|$return$|$ distribution looks
non-power-law because it averages distinct regimes is rejected
within the data: 2016--2020 and 2021--2026 each individually reject
at conventional levels. Only the smallest sub-period (2010--2015)
fails to reject, and as noted that is a tail-truncation issue
rather than evidence of a power-law fit.

\subsection{Comparison against alternatives via Vuong}
\label{sec:dist-vuong}

The bootstrap goodness-of-fit p-value reports whether the power
law is plausible against itself plus noise. Vuong's likelihood-ratio
test (Section~\ref{sec:vuong}) reports whether the power law is
preferred over alternative parametric distributions. The alternatives
considered here are those listed as step~4 of the CSN protocol in
Section~\ref{sec:csn-protocol}: \textit{lognormal}, \textit{exponential},
\textit{stretched exponential}, and \textit{power law with exponential
cutoff}. These are distributions for a random variable, distinct from
the trend specifications of Section~\ref{sec:candidates}, and are
the canonical alternatives in the CSN literature.
Across all eleven tests, Vuong selects \textbf{lognormal} as the
better fit with overwhelming evidence:
$R$ statistics range from approximately $-15$ on the smallest
sub-period (2010-2015, where the bootstrap was underpowered) to
$\sim\!-22{,}000$ on the largest UTXO snapshot, all with $p < 10^{-10}$.
(Negative $R$ indicates the second-listed model -- here, lognormal --
wins.) Where the bootstrap rejects PL, Vuong identifies lognormal
as the better fit. Where the bootstrap is underpowered, Vuong is
not, and the same lognormal preference holds.

The two procedures therefore agree, and the conclusion is robust to
which test one prefers: PL is not the right family for these tail
distributions; lognormal is.

\subsection{Conclusion of the distributional tests}
\label{sec:dist-conclusion}

In the distributional setting --- where the CSN protocol is
canonical and well-established --- Bitcoin's tail-relevant series do
not follow a power law. UTXO balances and daily $|$returns$|$ are
both better described by a lognormal in every test of sufficient
tail size, and the failures-to-reject are tail-truncation artefacts
rather than evidence in favour of a power law.

\paragraph{Series selection criteria.}
The distributional CSN protocol applies to random variables: a
cross-section of values, or the marginal distribution of an
approximately stationary series. UTXO balances at a snapshot
satisfy this naturally (one balance per UTXO, in a population of
millions). Daily $|r_t|$ is approximately stationary after first-differencing
the log-price series; the augmented Dickey--Fuller test
\citep{DickeyFuller1979, SaidDickey1984} rejects a unit root at
$p < 10^{-6}$ in each of the three sub-periods used below
($p = 5.9\!\times\!10^{-7}$, $6.9\!\times\!10^{-8}$,
$2.4\!\times\!10^{-7}$ for 2010-2015, 2016-2020, and 2021-2026
respectively, and $p = 1.7\!\times\!10^{-15}$ for the full sample). The
five on-chain time series we use elsewhere in the paper (hash rate,
unique addresses, transactions, difficulty, UTXO count) are scalar
series with strong long-run trends, and their marginal distributions
would conflate tail behaviour with growth dynamics rather than
isolate any genuine power-law tail. They are therefore not candidates
for the distributional test; we reserve them for the time-domain
power-law tests of Section~\ref{sec:time-domain} and the
wave-structure tests of Section~\ref{sec:waves}.

This sets the methodological backdrop for what follows. The
remainder of the paper turns to a separate power-law claim --- that
Bitcoin's price as a function of time follows $P \sim t^\beta$.
This is a regression statement rather than a distributional one
(Section~\ref{sec:csn-protocol}), and prior work that has tested
it has done so via OLS rather than via CSN
\citep{SantostasiPerrenod2026, Wheatley2019}. The CSN methodology
cannot be applied mechanically in this setting; the next section
develops principled diagnostic analogues and reports their
results.

\section{Time-domain power-law assessment}
\label{sec:time-domain}

Section~\ref{sec:distributional} applied the canonical CSN protocol to
Bitcoin's tail-relevant random variables and rejected the power-law
hypothesis in favour of a lognormal alternative. This section turns
to a different power-law claim: that Bitcoin's price as a function of
time follows $P \sim t^\beta$. As noted in
Section~\ref{sec:csn-protocol}, this is a regression statement rather
than a distributional one, so the CSN procedure does not apply
mechanically. We use the three adaptations developed in
Section~\ref{sec:csn-adaptations} (shift sensitivity, bootstrap
goodness-of-fit on residual diagnostics, Vuong's likelihood-ratio
test), supplement them with a parameter-matched flexibility check,
and reproduce standard scale-invariance tests proposed in earlier
work \citep{SantostasiPerrenod2026}.

The headline of this section is that the standard tests --- including
those proposed in earlier work --- demonstrate that Bitcoin's price
has a stable long-run average slope on log-log axes near
$\hat\alpha \approx 5.6$, but do not by themselves identify the
power law as the unique structural form generating the data. Several
alternative specifications, including a multi-component sigmoid
stack, are consistent with the same observations.
Section~\ref{sec:waves} introduces a new test that does
discriminate.

\subsection{Shift-parameter sensitivity (Adaptation A)}
\label{sec:s-sweep}

The time-domain regression is parameterised by a shift $s$,
\[
\log_{10}(P) \;=\; \alpha \, \log_{10}(t + s) + c,
\]
where $s = 0$ corresponds to the genesis block as the power-law
time origin (Section~\ref{sec:csn-adaptations}). An exponent that
is robust to specification should be approximately invariant to the
choice of $s$ over a reasonable range; large variation indicates
that the apparent law depends on a methodological choice and so
fails the most basic invariance any structural reading would
require. Table~\ref{tab:s-sweep} reports the sweep over
$s \in [0, 5000]$.

\begin{table}[htbp]
\centering
\caption{Shift-parameter sensitivity (Adaptation~A). For each $s$ we
refit the time-domain regression on Bitcoin's daily price
($n = 5{,}699$) and report the fitted $\hat\alpha$, $R^2$, AIC, and
residual standard deviation $\sigma$. AIC is minimised at $s = 0$.}
\label{tab:s-sweep}
\small
\begin{tabular}{rrrrr}
\toprule
$s$ (days) & $\hat\alpha$ & $R^2$ & AIC & $\sigma_{\text{resid}}$ \\
\midrule
0     & 5.65  & 0.9595 & \textbf{2{,}528} & 0.302 \\
30    & 5.73  & 0.9592 & 2{,}573 & 0.303 \\
60    & 5.81  & 0.9589 & 2{,}621 & 0.304 \\
120   & 5.97  & 0.9581 & 2{,}722 & 0.306 \\
240   & 6.28  & 0.9565 & 2{,}939 & 0.313 \\
365   & 6.59  & 0.9547 & 3{,}169 & 0.319 \\
500   & 6.92  & 0.9527 & 3{,}412 & 0.326 \\
1{,}000 & 8.09 & 0.9457 & 4{,}204 & 0.349 \\
2{,}000 & 10.29 & 0.9342 & 5{,}294 & 0.385 \\
5{,}000 & 16.49 & 0.9144 & 6{,}796 & 0.439 \\
\bottomrule
\end{tabular}
\end{table}

\begin{figure}[!t]
\centering
\includegraphics[width=0.85\textwidth]{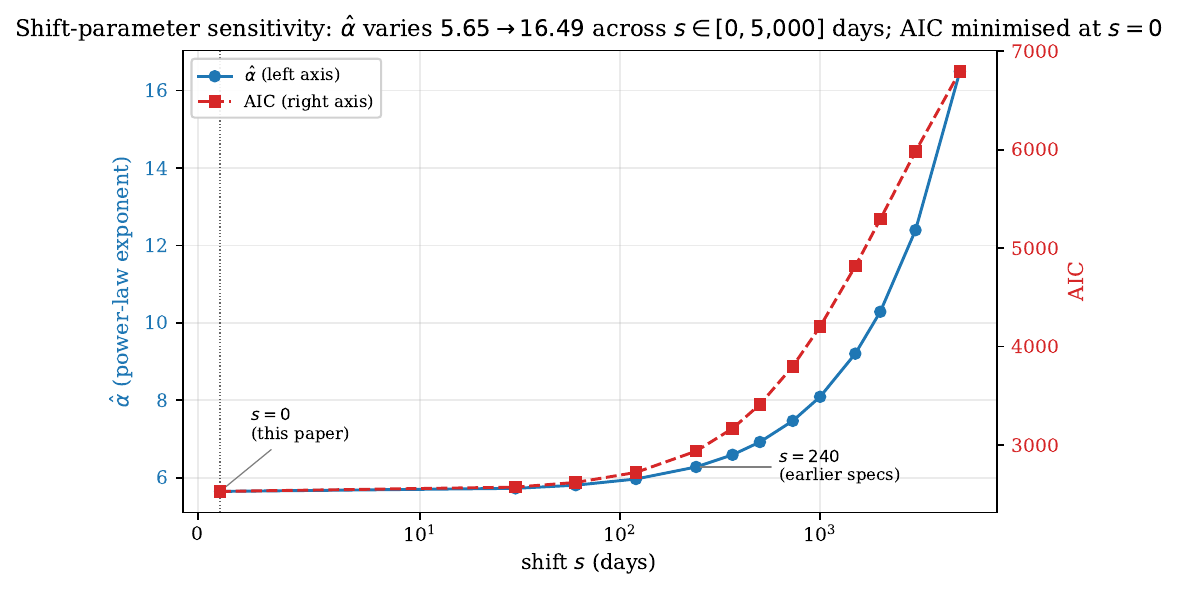}
\caption{Shift-parameter sensitivity. Fitted power-law exponent
$\hat\alpha$ (blue circles, left axis) and AIC (red squares,
right axis) as a function of the shift $s$ in
$\log_{10}(P) = \alpha\,\log_{10}(t + s) + c$. Both quantities rise
sharply and monotonically with $s$; AIC is minimised at $s = 0$
(our specification, also matching \citealt{SantostasiPerrenod2026}).
The fitted exponent varies by nearly a factor of three across the
swept range, indicating that $\hat\alpha$ is not a shift-invariant
structural property of the data.}
\label{fig:shift-sensitivity}
\end{figure}

The fitted $\hat\alpha$ varies by nearly a factor of three across
the range, from $5.65$ at $s = 0$ to $16.49$ at $s = 5{,}000$
(Figure~\ref{fig:shift-sensitivity}). By
AIC, $s = 0$ is the optimal specification --- the AIC penalty
grows monotonically with $s$, exceeding the AIC at $s = 5{,}000$
by over 4{,}000 units. We adopt $s = 0$ as the primary
specification throughout the rest of the paper. This matches
\citet{SantostasiPerrenod2026} and corresponds to placing the
power-law time origin at the Bitcoin genesis block.

The sweep is informative beyond the choice of specification: the
exponent $\hat\alpha$ is specification-dependent rather than
specification-robust --- it is a function of a free parameter $s$
that itself must be chosen, which is the most basic invariance a
structural-property reading would require.
The AIC-optimal value happens to coincide with the natural
reference point (the genesis block), but using a non-zero shift
of, e.g., $s = 240$ days --- a value motivated in
\citet{BaqueroTinoco2026} as a way to ``reduce leverage from
pre-2011 low-liquidity prices''\footnote{Their formulation places
$s$ relative to days-since-first-observation rather than
days-since-genesis, so the numerical equivalence is approximate;
the qualitative point that earlier work used non-zero shifts
holds.} --- one would report $\hat\alpha = 6.28$ rather than
$5.65$ in our coordinate. With $s = 1{,}000$ one would report
$\hat\alpha = 8.09$. Differences of this magnitude across ``the
same data'' reflect a specification-dependent exponent: there is no
single shift-invariant value that the procedure recovers from the
data alone.

\subsection{In-sample model comparison}
\label{sec:in-sample-results}

We compare the candidate trend specifications of
Section~\ref{sec:candidates} on the time-domain regression of
Bitcoin's daily price at $s = 0$. Table~\ref{tab:in-sample}
reports $R^2$, residual standard deviation, log-likelihood, AIC,
and BIC, sorted by AIC.

\begin{table}[htbp]
\centering
\caption{In-sample comparison of candidate trend specifications on
Bitcoin's daily price ($s = 0$, $n = 5{,}699$, $\log_{10}(P)$ as
the dependent variable). $\Delta$AIC is reported relative to the
power law. Negative $\Delta$AIC means the model is preferred by AIC.}
\label{tab:in-sample}
\small
\begin{tabular}{lrrrrrrr}
\toprule
Model & params & $R^2$ & $\sigma$ & log-lik & AIC & BIC & $\Delta$AIC \\
\midrule
Multi-sigmoid (K=3)     & 10 & 0.9742 & 0.241 & $+19.3$  & $-19$    & $48$    & $-2{,}547$ \\
Polynomial (deg 9)      & 10 & 0.9658 & 0.278 & $-783.7$ & $1{,}587$ & $1{,}654$ & $-941$ \\
Cubic B-spline (10)     & 10 & 0.9648 & 0.282 & $-864.5$ & $1{,}749$ & $1{,}815$ & $-779$ \\
\textbf{Power law}      & 2  & 0.9595 & 0.302 & $-1{,}262$ & $\mathbf{2{,}528}$ & $2{,}541$ & 0 \\
Stretched exponential   & 3  & 0.9595 & 0.302 & $-1{,}262$ & $2{,}530$ & $2{,}550$ & $+2$ \\
Cubic                   & 4  & 0.9559 & 0.315 & $-1{,}509$ & $3{,}025$ & $3{,}052$ & $+497$ \\
Sigmoid (K=1)           & 4  & 0.9522 & 0.328 & $-1{,}739$ & $3{,}485$ & $3{,}512$ & $+957$ \\
Quadratic               & 3  & 0.9474 & 0.344 & $-2{,}009$ & $4{,}023$ & $4{,}043$ & $+1{,}495$ \\
Pure exponential        & 2  & 0.8738 & 0.533 & $-4{,}502$ & $9{,}008$ & $9{,}021$ & $+6{,}480$ \\
\bottomrule
\end{tabular}
\end{table}

Four observations follow.

\paragraph{1. Smooth alternatives collapse to the power law.}
The stretched exponential ranks alongside the power law: identical
$R^2$ to four decimal places, identical $\sigma$, and a $\Delta$AIC
of exactly $+2$ (the parameter penalty for the unused $\beta$).
The optimiser drives the stretched-exponential's $\beta$ parameter
to $\sim 10^{-6}$, in which limit
\(t^\beta = e^{\beta \ln t} \approx 1 + \beta \ln t\), and the
model algebraically reduces to a power law in $\ln t$ with effective
exponent $b\beta$. Computing $b\beta \cdot \ln 10$ recovers the
directly-fitted $\hat\alpha = 5.644$ to five decimal places. This
collapse, observed informally in an earlier public
discussion,\footnote{\url{https://x.com/Giovann35084111/status/2049786216988737700}}
indicates that the \emph{smooth-trend class} in our candidate set
is essentially equivalent to the power law on this data.

\paragraph{2. The pure exponential is decisively rejected.}
$\Delta$AIC $= +6{,}480$ relative to the power law. Bitcoin does
not grow exponentially over our window; the casual ``exponential
growth'' framing is inconsistent with the data.

\paragraph{3. The multi-sigmoid stack ($K = 3$) wins by a
substantial margin in-sample.}
$\Delta$AIC $= -2{,}547$; Vuong's BIC-corrected likelihood-ratio
test (Section~\ref{sec:vuong}) gives $V = -19.74$, $p < 10^{-80}$.
The result is robust to the choice of complexity penalty: BIC,
which penalises the multi-sigmoid's eight extra parameters more
heavily than AIC, also prefers $K = 3$ over the power law by
$\Delta\mathrm{BIC} = 48 - 2{,}541 = -2{,}493$. Notably, the single
sigmoid ($K = 1$) is substantially worse than the power law
($\Delta$AIC $= +957$, $\Delta$BIC $= +971$), so the in-sample
improvement of $K = 3$ comes from the multi-component structure,
not from a single saturation event.

\paragraph{4. Parameter-matched flexibility check.}
Two flexible alternatives --- polynomial of degree 9 (10
parameters, in normalised $t$) and cubic B-spline with 10 basis
functions --- match $K = 3$ in parameter count. Both improve over
the power law substantially ($\Delta$AIC $= -941$ and $-779$
respectively), but $K = 3$ still wins by an additional 1{,}605
and 1{,}767 AIC units beyond the flexibility budget. This indicates
that the in-sample advantage of $K = 3$ is not driven solely by
having more parameters: the sigmoid functional form contributes
beyond raw flexibility.

\paragraph{Caveat: AIC is not a sufficient structural test.}
The raw $\Delta$AIC of $-2{,}547$ for $K = 3$ vs the power law
overstates the structural evidence. As a control, we generated a
synthetic dataset of the same length under the null that the data
is a true power law plus AR(1) noise (with $\rho$ and $\sigma$
calibrated to Bitcoin's residuals); the same comparison on this
synthetic gives $\Delta$AIC $= -4{,}552$ for $K = 3$ vs PL ---
larger in magnitude than what we observe on real Bitcoin.
Strong AR(1) noise produces slow-varying residual patterns that
the multi-sigmoid's flexibility can absorb, inflating the AIC
advantage of the more flexible model regardless of the underlying
generative process. AIC alone therefore cannot distinguish a true
power-law generator from a multi-component generator on data of
this length and autocorrelation. For our in-sample structural
evidence, we rely on the parameter-matched flexibility check
(point 4 above), which is unaffected by this caveat: all three
ten-parameter alternatives (multi-sigmoid, polynomial, B-spline)
share the same parameter budget, so the AIC gap among them
isolates the contribution of functional form.

\subsection{Bootstrap goodness-of-fit on residual diagnostics
(Adaptation~B)}
\label{sec:bootstrap-gof}

The power-law residuals on real Bitcoin price exhibit several
pathologies at conventional significance levels. We use five
standard residual tests, each defined briefly here for the reader
who has not seen them before. In each test the null hypothesis is
that residuals behave as expected under a well-specified
regression with independent, identically distributed Gaussian
noise; a small p-value rejects the null and indicates a specific
form of misspecification.

\begin{itemize}[leftmargin=2em]
\item \emph{Ljung--Box test \citep{LjungBox1978}.} Tests whether
residuals at lags $1, 2, \ldots, h$ are jointly uncorrelated.
Rejection indicates serial autocorrelation. On real BTC at
$h = 50$, the test statistic is $\sim\!2.4\!\times\!10^{5}$ with
$p \approx 0$.
\item \emph{Breusch--Pagan test \citep{BreuschPagan1979}.} Tests
whether residual variance depends on the regressors. Rejection
indicates heteroscedasticity. On real BTC the LM statistic is
$\sim\!822$ with $p < 10^{-180}$.
\item \emph{Shapiro--Wilk test \citep{ShapiroWilk1965}.} Tests
whether residuals are normally distributed; rejection indicates
non-normality. On a 5{,}000-point subsample of BTC residuals
(the maximum sample the test handles tractably), $p < 10^{-40}$.
\item \emph{Jarque--Bera test \citep{JarqueBera1980}.} A second
normality test based on skewness and excess kurtosis of the
residuals. On real BTC the statistic is $\sim\!693$, $p < 10^{-150}$,
with skewness $0.83$ and excess kurtosis $0.41$.
\item \emph{Chow break test \citep{Chow1960}.} Tests whether the
regression coefficients are stable across a candidate breakpoint;
rejection indicates a structural break. On real BTC the test is
applied at six biennial candidate dates from 2013 to 2023; all
six reject at conventional significance.
\end{itemize}

These rejections are conventional in the sense that the residuals
are clearly not the well-behaved Gaussian white noise of a
correctly specified regression. Adaptation~B asks whether they are
\emph{discriminating}: would they distinguish a power law plus
realistic noise from data generated by some alternative process?

We generate $B = 100$ synthetic trajectories under the null that
the data is generated by the fitted power law plus AR(1) noise
(Section~\ref{sec:bootstrap}; calibrated $\hat\rho = 0.998$,
$\hat\sigma = 0.302$). For each synthetic trajectory we compute
the same residual diagnostics. The bootstrap p-value is the
fraction of synthetic trajectories whose diagnostic value
matches or exceeds the observed.

The result is uncomfortable but informative: across all five
diagnostics, the synthetic distribution either matches or exceeds
the observed values. Bootstrap p-values exceed 0.50 for each
diagnostic; in over half of the synthetic trajectories generated
from a true power law plus realistic noise, the residuals show
``pathology'' of the same magnitude as observed on real Bitcoin.

The conclusion is methodologically important: standard residual
diagnostics, applied to a power-law regression on a series with
the autocorrelation structure of Bitcoin's price, are not
informative for distinguishing PL plus noise from alternative
trend models. They reject the null of well-behaved residuals,
but they do so equally on data that genuinely is a power law
plus realistic noise. Such diagnostics cannot serve as evidence
against the power-law specification --- a lesson that applies to
any test of power-law claims on heavily autocorrelated time
series.

\subsection{Reproduction of scale-invariance tests}
\label{sec:scale-invariance}

A separate line of argument for the power-law specification rests
on tests of scale invariance --- properties of the price--time
relationship that should be preserved if the underlying process
is truly $P \sim t^\beta$. Earlier work \citep{SantostasiPerrenod2026}
proposes four such tests on Bitcoin's price: a pair-ratio test, a
direct collapse test, a rolling-window stability test, and a
Bayesian sequential update on local-slope estimates. We reproduce
all four on Bitcoin's price series and on a synthetic comparison
trajectory: a 3-component sigmoid stack fit to the same prices,
evaluated at the same time points.

The rationale for the synthetic comparison is that
scale-invariance tests are typically motivated by the question
``does the data behave like a power law?'' --- a binary check
against an unspecified alternative. The 3-sigmoid synthetic, fit
to match the actual price trajectory, has the same long-run
average slope by construction; if the tests cannot distinguish it
from the real data, they are passing on a property that does not
identify the power law specifically.

\paragraph{Pair-ratio test.}
For 4{,}000 random pairs of time points $(t_1, t_2)$ with
$t_2 > t_1$, compute $\log_{10}(P(t_2)/P(t_1))$ and plot against
$\log_{10}(t_2 / t_1)$. A power law $P(t) = A t^\beta$ produces a
straight line with slope $\beta$. We replicate the test on real
Bitcoin and on the sigmoid synthetic.

\begin{center}
\begin{tabular}{lcc}
\toprule
& Real BTC & 3-sigmoid synthetic \\
\midrule
Binned slope & 5.69 & 5.74 \\
$R^2$ & 0.998 & 0.999 \\
Quadratic curvature & 0.42 & 0.33 \\
\bottomrule
\end{tabular}
\end{center}

Both produce essentially perfect linearity with closely matched
slopes. The synthetic actually outperforms real Bitcoin slightly
(higher $R^2$, smaller curvature), since it has no noise. The
test does not distinguish.

\paragraph{Direct collapse test.}
Find the exponent $\beta^*$ that makes the residual mean of
$\log_{10}(P(\lambda t)/P(t)) - \beta \log_{10}\lambda$ flat as a
function of $\log_{10}\lambda$ for $\lambda \in [1.1, 5.0]$. Under
a true power law, $\beta^* = \beta$.

\begin{center}
\begin{tabular}{lcc}
\toprule
& Real BTC & 3-sigmoid synthetic \\
\midrule
OLS $\beta$ & 5.62 & 5.65 \\
Collapse $\beta^*$ & 5.60 & 5.64 \\
\bottomrule
\end{tabular}
\end{center}

Both yield consistent $\beta^*$ values that match their direct
OLS estimates. The test does not distinguish.

\paragraph{Rolling temporal stability.}
A rolling-window estimate $\beta^*(t)$ should be stationary if
the underlying generative process is a stable power law. We
compute rolling estimates over 145 windows.

\begin{center}
\begin{tabular}{lcc}
\toprule
& Real BTC & 3-sigmoid synthetic \\
\midrule
Median $\beta^*$ & 5.74 & 5.79 \\
Std $\beta^*$ & 0.48 & 0.37 \\
Drift slope (per day) & $1.7\!\times\!10^{-5}$ ($p=0.49$) & $2.3\!\times\!10^{-5}$ ($p=0.23$) \\
\% above median & 62\% & 74\% \\
\bottomrule
\end{tabular}
\end{center}

Neither shows significant secular drift; both are quasi-stationary
with similar median and amplitude. The test does not distinguish.

\paragraph{Bayesian sequential update.}
A conjugate Gaussian update on local $\hat\beta$ estimates should
shrink the posterior precision $\tau_n$ as $\sigma_{\text{emp}}/\sqrt n$
if the underlying parameter is constant. We compute this for both
datasets, with the prior calibrated separately to each
dataset's empirical local-estimate variance. The resulting
$\tau_n / (\sigma_{\text{emp}}/\sqrt n)$ is essentially 1 in both
cases (real BTC: $0.999 \pm 0.001$; synthetic: $1.000 \pm 0.001$,
across the full sequence). Both datasets exhibit textbook
$1/\sqrt n$ shrinkage of the posterior precision. The test does
not distinguish.

\paragraph{What the four tests collectively demonstrate.}
The four scale-invariance tests confirm that Bitcoin's price has a
stable long-run average slope near $\hat\alpha \approx 5.6$ over
the observation window. This is a real and remarkable empirical
regularity. But the tests are constructed to detect departures
from a stable average slope; they pass as soon as such an average
exists, regardless of whether the underlying generative process
is a true power law or a multi-component model whose envelope
matches. The 3-sigmoid synthetic, fit to match Bitcoin's
trajectory, has the same long-run average slope by construction,
and consequently passes the same tests. The tests are necessary
but not sufficient for a structural power-law claim.

\subsection{Conclusion of the time-domain assessment}
\label{sec:time-domain-conclusion}

Four results from this section can be summarised as follows.

\begin{itemize}[leftmargin=2em]
\item The power-law exponent $\hat\alpha$ depends on the shift
parameter $s$, varying by nearly a factor of three across the
range $s \in [0, 5{,}000]$. The fitted exponent is specification-dependent
rather than specification-robust --- it is not a shift-invariant
property of the data alone.
\item The multi-sigmoid stack ($K = 3$) wins the in-sample
comparison decisively in AIC and Vuong's likelihood-ratio test.
The win persists at parameter-matched count against polynomial
and B-spline alternatives, indicating that the sigmoid functional
form contributes beyond raw flexibility, although the raw AIC
advantage is partially inflated by AR(1) noise.
\item Standard residual diagnostics on the power-law fit reject
at conventional levels, but they reject equally on synthetic data
generated as a true power law plus realistic noise. They are not
discriminating.
\item Scale-invariance tests proposed in earlier work pass on
real Bitcoin, but they also pass on a 3-sigmoid synthetic fit to
the same data. They demonstrate the long-run average slope is
stable, which is real, but they do not by themselves identify the
power law as the unique structural form.
\end{itemize}

These results establish two things. First, in the canonical sense
required for a structural reading --- a fitted exponent invariant to
the basic specification choices --- Bitcoin's price-time relationship
fails the test: the exponent depends on the shift parameter, and the
standard residual and scale-invariance tests cannot distinguish a
power law from a multi-component alternative on this data. Second, the in-sample data contains
genuine information that prefers the multi-component alternative,
but the standard distributional and scale-invariance tests
cannot extract it. What the data actually says about Bitcoin's
price structure --- and how to test it rigorously --- is the
subject of the next section.

\section{Wave structure: a Bitcoin-specific finding}
\label{sec:waves}

Section~\ref{sec:time-domain} established that the standard tests of
structural form --- CSN adaptations and the scale-invariance tests
proposed in earlier work --- cannot distinguish a power law from
alternative trend specifications on Bitcoin's price. The data
contains genuine information that prefers a multi-component
description, but the standard tests do not extract it.

This section introduces a test that does discriminate, using
cross-series comparison: the in-sample functional-form comparison
runs across nine series (Bitcoin price, five Bitcoin on-chain
metrics, and three major traditional asset classes); the
quarterly $K = 3$ wave-stability bootstrap runs on a seven-series
subset (Bitcoin price, hash rate, NASDAQ, S\&P~500, gold, plus
Ethereum and a lithium ETF as additional candidate adoption-driven
assets). Bitcoin price is the only series in the in-sample test
where no single-component growth model improves on the power law.
In the wave-stability bootstrap, Bitcoin price rejects the PL+AR(1)
null at $p_{<15\%} = 0.015$; the other six series sit in the null
(Ethereum is the second-closest to rejection at
$p_{<25\%} = 0.030$, the rest comfortably above conventional
thresholds).

\subsection{\texorpdfstring{Functional form: $K = 1$ sigmoid versus power law across nine series}{Functional form: K=1 sigmoid versus power law across nine series}}
\label{sec:k1-asymmetry}

We applied the in-sample model comparison from
Section~\ref{sec:in-sample-results} to nine series, all over the
same calendar window 2010-08-18 to 2026-03-25 (the on-chain
series end one to two days earlier, at 2026-03-16). For each series we
fit the candidates of Section~\ref{sec:candidates} and recorded
the AIC difference relative to the power law. Table~\ref{tab:k1-pl}
focuses on the comparison most relevant to identifying Bitcoin
price's distinctive behaviour: the single-sigmoid (K=1)
specification against the power law.

\begin{table}[htbp]
\centering
\caption{Single-component model comparison across nine series. For
each series we report the $\Delta$AIC of the K=1 sigmoid and of
the pure exponential, both relative to the power law on the same
series. Negative values indicate the alternative is preferred to
the power law. \textbf{Bitcoin price is the only series in the
sample where the K=1 sigmoid does not improve on the power law},
and the only series where neither single-component alternative
(K=1 sigmoid or pure exponential) wins.}
\label{tab:k1-pl}
\small
\begin{tabular}{lrrl}
\toprule
Series & $\Delta$AIC: K=1 vs PL & $\Delta$AIC: Pure exp vs PL & K=1 wins? \\
\midrule
\textbf{Bitcoin price}             & \textbf{$+957$}  & $+6{,}480$ & \textbf{No (PL wins)} \\
Bitcoin hash rate                  & $-4{,}781$       & $+8{,}784$ & Yes \\
Bitcoin unique addresses           & $-9{,}123$       & $+4{,}908$ & Yes \\
Bitcoin transactions               & $-6{,}022$       & $+5{,}055$ & Yes \\
Bitcoin difficulty                 & $-4{,}797$       & $+8{,}845$ & Yes \\
Bitcoin UTXO count                 & $-5{,}164$       & $+7{,}617$ & Yes \\
\midrule
NASDAQ Composite                   & $-5{,}265$       & $-5{,}263$ & Yes (\textit{$\approx$ exp}) \\
S\&P 500                           & $-5{,}706$       & $-5{,}768$ & Yes (\textit{$\approx$ exp}) \\
Gold (continuous front-month)      & $-6{,}198$       & $-1{,}340$ & Yes \\
\bottomrule
\end{tabular}
\end{table}

\paragraph{Three regimes visible across the nine series.}
The same K=1 specification behaves differently across the nine
series, and the pattern is informative.

For \textbf{NASDAQ and S\&P 500}, the K=1 sigmoid and the pure
exponential perform almost identically (within 60 AIC units of each
other), and both beat the power law by $\sim$5{,}000-6{,}000 AIC
units. Mathematically this is expected: for $t$ well below a
sigmoid's inflection point, $\log_{10}(P) = b + L/(1 + e^{-k(t-t_0)})$
is approximately linear in $t$, so the sigmoid \emph{degenerates}
to a pure exponential. For these two indices the saturation point
$t_0$ falls past 2026, and the observed window lies entirely in
the early-exponential phase of any fitted single sigmoid. Both
specifications therefore describe the same empirical fact:
NASDAQ and S\&P~500 grew approximately exponentially over our
window.

For \textbf{gold}, the K=1 sigmoid wins by 4{,}858 more AIC units
than the pure exponential. Gold has visible within-window
dynamics that an exponential cannot capture (a rise to ${\sim}\$1{,}900$
in 2011, plateau and decline through 2018, rise again 2019+). The
sigmoid's saturation curve fits this; the exponential cannot.

For \textbf{the five Bitcoin on-chain metrics}, the K=1 sigmoid
beats the power law by 4{,}800-9{,}100 AIC units, and decisively
beats the pure exponential. The on-chain series are well described
as smooth saturation curves --- adoption-curve shapes consistent
with the diffusion-of-innovations literature
\citep[][also Section~\ref{sec:candidates}]{Bass1969}.

For \textbf{Bitcoin price}, neither specification works:
$\Delta$AIC $= +957$ for K=1 vs the power law, and $+6{,}480$ for
pure exponential. \textbf{It is the only series in our nine-series
sample where no single growth component --- whether exponential,
saturating, or otherwise --- improves on the power law.} The
multi-component K=3 specification, in contrast, beats the power
law on Bitcoin price by 2{,}547 AIC units (Table~\ref{tab:in-sample}).
The result holds at parameter-matched count
(Section~\ref{sec:in-sample-results}, observation~4).

The K=1-versus-PL asymmetry is therefore not a result internal
to Bitcoin's on-chain ecosystem, nor a feature of the comparison
methodology. It is a property specific to Bitcoin price, and not
to any of the eight other series tested.

\begin{figure}[!t]
\centering
\includegraphics[width=\textwidth]{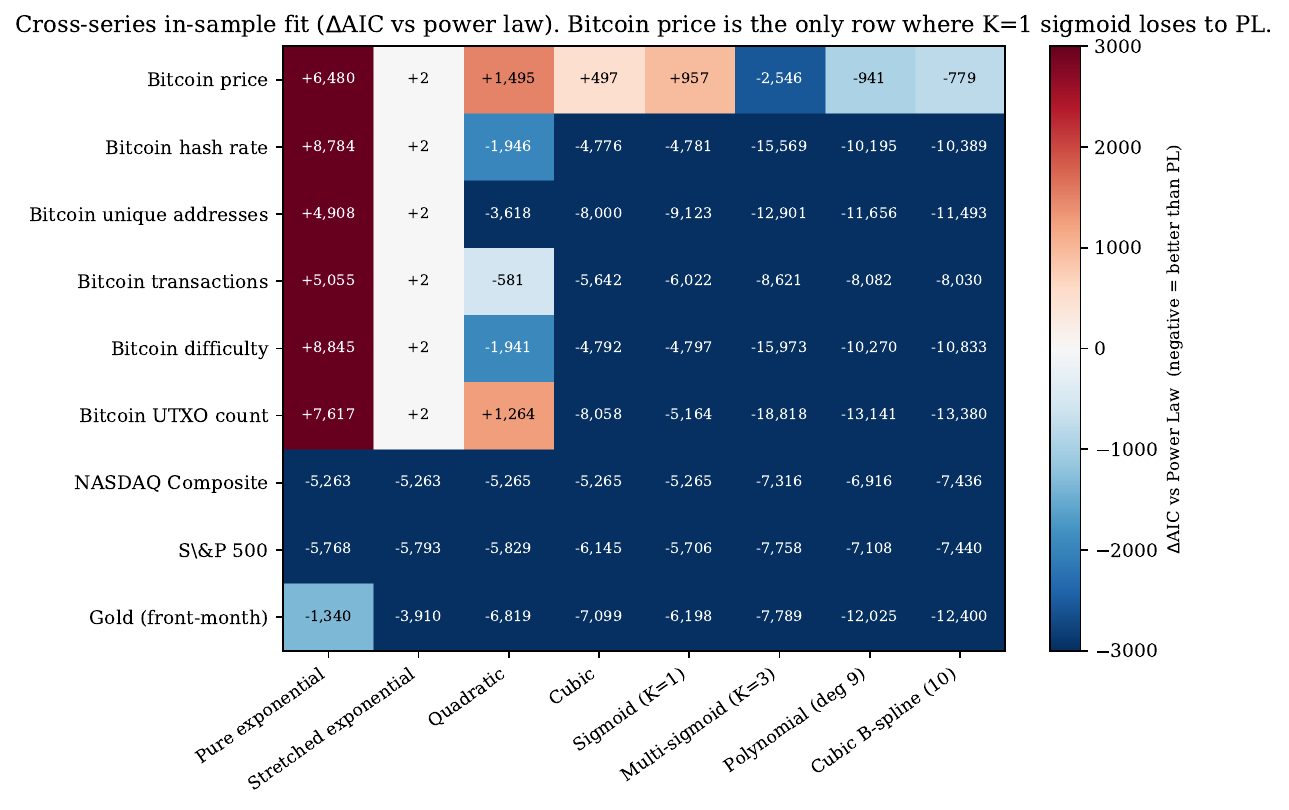}
\caption{Cross-series in-sample fit comparison. For each of the
nine series (rows) and each of the eight non-power-law candidates
(columns), the cell shows $\Delta$AIC vs the power-law fit on the
same series. Negative (blue) values mean the candidate beats the
power law; positive (red) values mean the power law wins. Bitcoin
price (top row) is the only row where both single-component
candidates --- the K=1 sigmoid and the pure exponential --- have
positive $\Delta$AIC. Stretched exponential collapses to the
power law on every series ($\Delta$AIC = $+2$ exactly,
the parameter-penalty cost of the unused $\beta$).}
\label{fig:cross-aic}
\end{figure}

Figure~\ref{fig:cross-aic} shows the full $\Delta$AIC pattern
across all eight non-power-law candidates and all nine series. The
Bitcoin price row is the only row in which the $K = 1$ and Pure
exponential cells are positive (worse than power law); every other
series has at least one single-component model that improves on
the power law.

\subsection{\texorpdfstring{The $K = 3$ wave-stability bootstrap}{The K=3 wave-stability bootstrap}}
\label{sec:wave-stability}

The single-component-fails finding suggests Bitcoin price has
multi-component structure. The K=3 wave-stability bootstrap
formalises this and tests it against the null hypothesis that the
data is a power law plus realistic noise.

\begin{figure}[!t]
\centering
\includegraphics[width=0.85\textwidth]{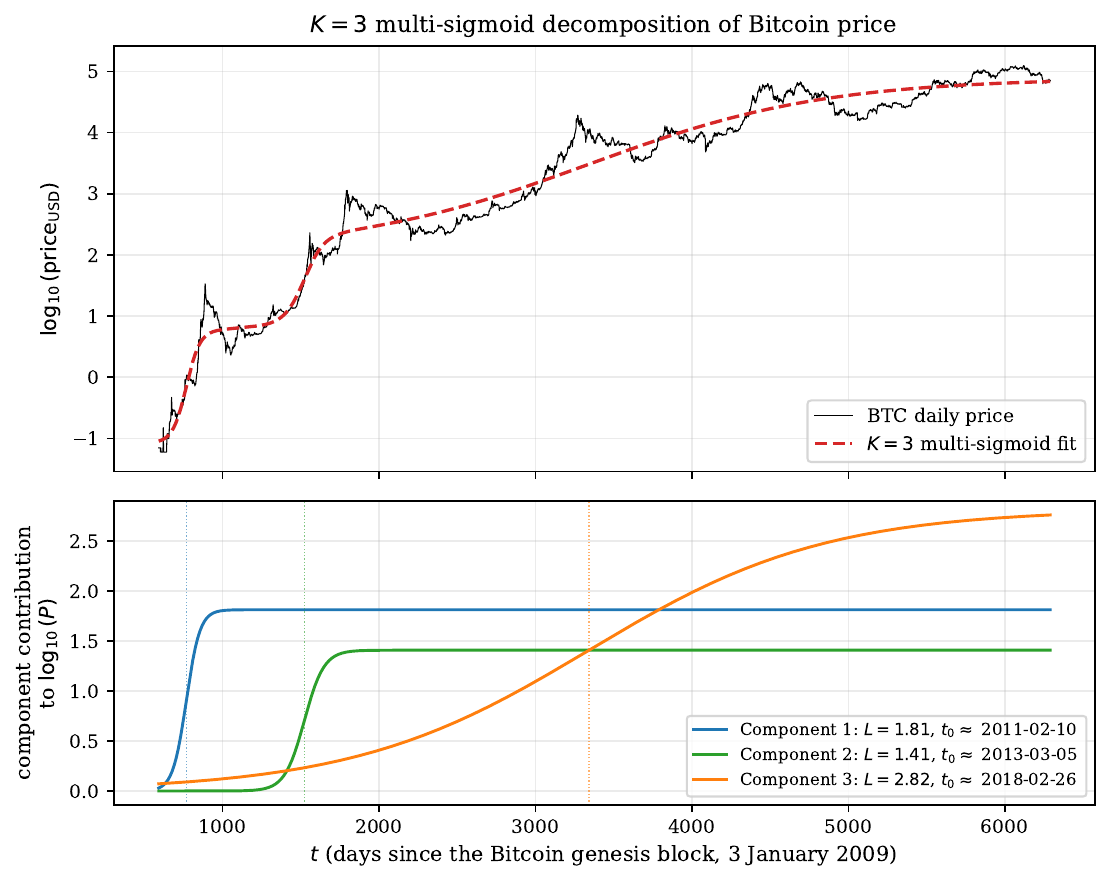}
\caption{Top: Bitcoin daily price (black) and the multi-sigmoid
($K = 3$) fit on the full sample (red dashed). Bottom: the three
fitted sigmoid components individually, sorted by inflection
date. Each component contributes a saturating S-curve; their sum
plus a baseline equals the top-panel fit. Inflection dates
$t_0$ are labeled at the foot of each sigmoid for reference.
The wave-stability bootstrap of
Section~\ref{sec:wave-stability} tests whether the amplitudes
$L$ of these three components are identifiable consistently when
the same $K = 3$ fit is repeated at successive rolling cutoffs.}
\label{fig:k3-decomposition}
\end{figure}

\paragraph{The test.}
The K=3 multi-sigmoid (Figure~\ref{fig:k3-decomposition}, top
panel) is fit at 41 quarterly cutoff dates
(2016-Q1 through 2026-Q1); for each cutoff we record the
amplitude $L_i$ of each of the three component sigmoids. A
component is classified as \emph{stable} if its amplitude has
coefficient of variation $\mathrm{CV} < 25\%$ across the
41 cutoffs.\footnote{We also report the stricter threshold
$\mathrm{CV} < 15\%$. The 25\% threshold is permissive enough
that a marginally identifiable wave can clear it; the 15\%
threshold demands very tight stability.} We count the number of
stable components on the real data.

\paragraph{Amplitude as the test statistic.}
Each fitted sigmoid component has three parameters --- amplitude
$L_i$, slope $k_i$, and inflection date $t_{0,i}$ --- and any of
them could in principle anchor a stability test. We choose
amplitude for four reasons. First, $\mathrm{CV} = \sigma/\mu$ is
dimensionless, so a single threshold (25\%, 15\%) applies
uniformly across series whose amplitudes differ by orders of
magnitude. Inflection dates are measured in days, and a
``stable'' threshold for them would require ad-hoc normalisation
relative to the training window. Second, amplitude has direct
interpretation as the persistent log-price gain attributable to a
wave; inflection dates depend on the phase of the wave and are
less interpretable when the wave is still in progress. Third,
amplitudes are bounded below by construction
($L \geq 0$), while inflection dates can saturate at the boundary
of the training window in early cutoffs and produce noisy
variance estimates. Fourth, under the PL+AR(1) null,
$L$ wanders unboundedly across spurious K=3 fits, whereas $t_0$
is constrained to lie inside the data range; the wider
null distribution on $L$ gives the test more discrimination
power against a stably-identified real wave.

\paragraph{Choice of $K = 3$.}
The multi-sigmoid family
(Section~\ref{sec:candidates}) admits any positive integer $K$;
the choice of $K = 3$ is motivated as the smallest stack that
contains the wave count visible in Bitcoin's price history over
our window: two clearly completed cycles (2011-2013, 2015-2017)
and one in progress (2020-2024). Smaller $K$ would force the fit
to merge cycles; larger $K$ would introduce components for
hypothetical waves not yet observed and could not be stably
identified. A $K = 4$ or $K = 5$ stability bootstrap would test
whether additional waves are statistically detectable, and is a
natural extension of the test below; we do not pursue it here.

If Bitcoin price has genuine multi-component structure, the K=3
fits at successive cutoffs should consistently identify the same
underlying components, and at least some component amplitudes
should be stable across cutoffs. If Bitcoin price is just a
power law with autocorrelated noise, the K=3 fit will overfit
different patterns at different cutoffs, and component amplitudes
will be unstable.

To test, we generate $B = 200$ synthetic trajectories under the
null that the data is a power law plus AR(1) noise (calibrated
$\hat\rho = 0.998$, $\hat\sigma = 0.302$, matching Bitcoin's own
residuals --- Section~\ref{sec:bootstrap}). For each synthetic
trajectory we re-run the K=3 fits at the same 41 cutoffs and
count stable components. The bootstrap p-value is the fraction
of synthetic trajectories that produce at least as many stable
components as the real data.

\paragraph{Conservatism of the test.}
A negative-control simulation, in which the synthetic generator
is instead a true 3-sigmoid stack plus the same AR(1) noise,
yields a wave-stability bootstrap p-value of approximately $0.50$
in our calibration. The test is biased toward null results: even
on data that genuinely has 3-component structure, the bootstrap
often fails to detect it. Reasons include: at early cutoffs (e.g.,
2016) the K=3 fit must accommodate three sigmoids over only one
or two empirically-visible waves, producing variable component
parameters by construction. We mention this here because it
means \emph{small} p-values from this test are conservative
indicators of true wave structure, not anti-conservative.

\begin{figure}[!t]
\centering
\includegraphics[width=\textwidth]{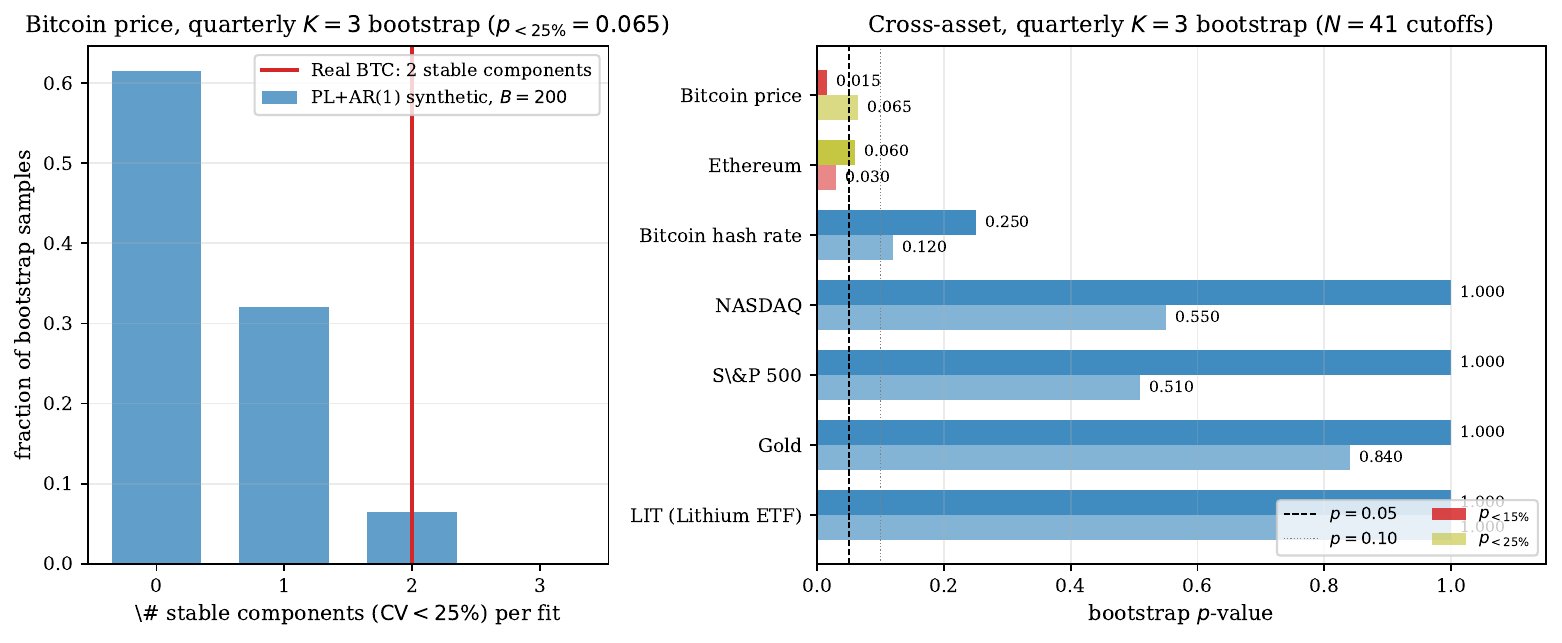}
\caption{Quarterly $K = 3$ wave-stability bootstrap. Left: the
bootstrap distribution for Bitcoin price. Bars give the fraction
of $B = 200$ synthetic PL+AR(1) trajectories producing 0, 1, 2,
or 3 stable components at the loose $\mathrm{CV} < 25\%$
threshold. The vertical red line marks the real Bitcoin
observation (2 stable components); only $13/200 = 6.5\%$ of
synthetic trajectories produce as many or more, giving the
bootstrap $p_{<25\%} = 0.065$. The corresponding strict-threshold
test gives $p_{<15\%} = 0.015$ (Bitcoin row of the right panel).
Right: bootstrap p-values for the same quarterly test applied to
the seven-series cross-asset comparison
(Table~\ref{tab:cross-asset}); both the strict (15\%) and loose
(25\%) thresholds are reported per series. Red bars are below
$p = 0.05$; tan bars are between $0.05$ and $0.10$; blue bars are
above $0.10$. Bitcoin and Ethereum are the only two series with
any p-value below $0.05$.}
\label{fig:wave-bootstrap}
\end{figure}

\paragraph{Result for Bitcoin price.}
We adopt quarterly cutoffs ($N = 41$ from 2016-Q1 through
2026-Q1) as the primary resolution: this is the coarsest
schedule for which the per-component CV estimator has
acceptable variance, and the finest schedule for which adjacent
cutoffs share meaningfully new training data.\footnote{A
lower-resolution yearly schedule ($N = 11$) gives directionally
consistent results with weaker statistical power
($p_{<15\%}^{\text{yearly}} = 0.190$,
$p_{<25\%}^{\text{yearly}} = 0.055$); the strengthening at the
strict threshold under the quarterly schedule reflects the more
accurate per-component CV estimate at higher $N$.} Real Bitcoin
price produces 2 stable components at the strict 15\% CV
threshold and 2 stable components at the 25\% threshold. Of 200
synthetic PL+AR(1) trajectories, 3 produce $\geq 2$ stable
components at 15\% and 13 at 25\%, giving the bootstrap p-values
$p_{<15\%} = 0.015$ and $p_{<25\%} = 0.065$
(Figure~\ref{fig:wave-bootstrap}, left panel).

\paragraph{Corroboration from inflection-date stability.}
As a corroborating check, we also record the standard deviation
of each component's inflection date $t_{0,i}$ across the 41
cutoffs. On real Bitcoin price these are $[19, 149, 505]$~days
for waves 1, 2, 3 respectively: the first wave's inflection is
pinned to within $\sim 19$~days across a decade of rolling
cutoffs, the second to within five months, while the third (the
2020--24 cycle, still in progress at the end of our window) has
a much wider posterior. The pattern matches the amplitude-CV
ranking ($7.2\%, 14.1\%, 36.0\%$): the same two waves identified
as amplitude-stable also have tightly identified inflection
dates. We report this as a supplementary check rather than as a
primary test, for the reasons given above (no natural threshold;
boundary effects at early cutoffs); the numerical values are
saved in the bootstrap JSON outputs.

The single-panel parallel-coordinates view in
Figure~\ref{fig:wave-bootstrap-quarterly} shows all 200
synthetic realizations of the per-wave CV vector against the
real-data realization, making the per-component discrimination
visible:

\begin{figure}[!t]
\centering
\includegraphics[width=0.85\textwidth]{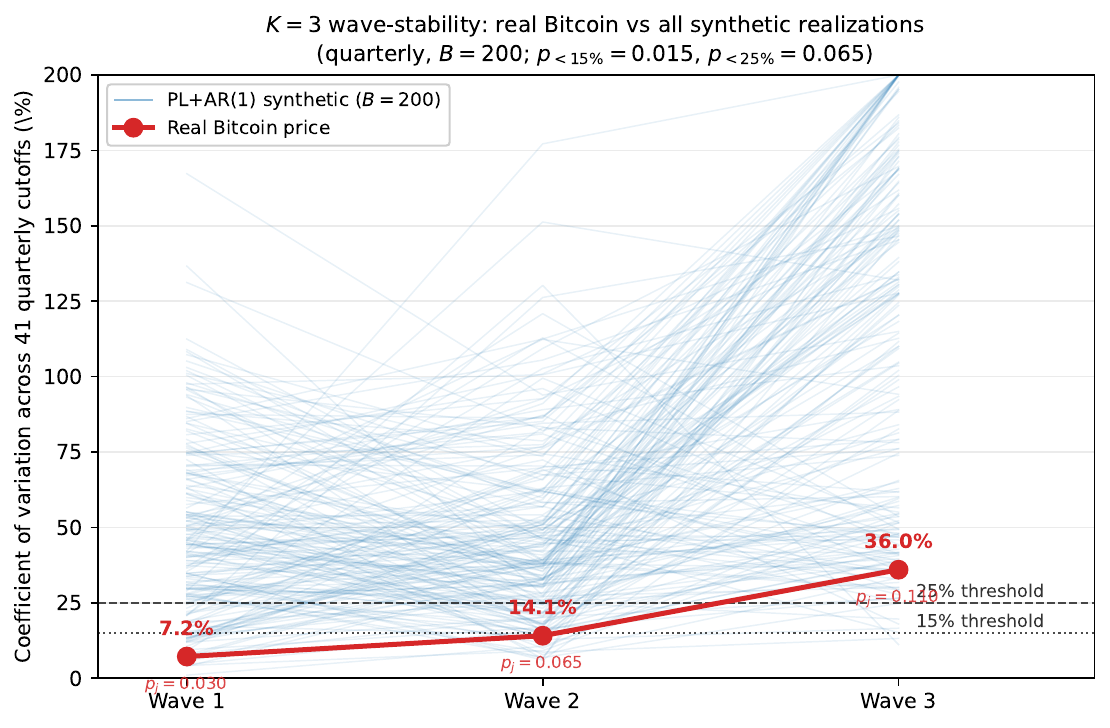}
\caption{$K = 3$ wave-stability bootstrap on Bitcoin price at
quarterly resolution ($N = 41$ cutoffs, $B = 200$). Each thin
blue line is one PL+AR(1) synthetic realization plotted as
$(\mathrm{CV}_1, \mathrm{CV}_2, \mathrm{CV}_3)$ across the three
sigmoid components ordered by inflection date. The thick red line
is the real Bitcoin observation, with per-wave empirical
p-values $p_j = \#\{\mathrm{synth} \leq \mathrm{real}\}/B$ shown
beneath each marker. Real Bitcoin's CV vector sits in the lower
tail of the synthetic cloud at every component, and below the
15\% threshold for waves 1 and 2 --- the source of the
$p_{<15\%} = 0.015$ rejection. By contrast, individual synthetic
realizations rarely produce two CVs below 15\% jointly, which is
the property the test discriminates on.}
\label{fig:wave-bootstrap-quarterly}
\end{figure}

\paragraph{Interpretation.}
At quarterly resolution, the strict-threshold result rejects
the PL+AR(1) null at conventional significance
($p_{<15\%} = 0.015$); the loose-threshold result is marginal
($p_{<25\%} = 0.065$, just above $0.05$ and detectable at the
10\% level). Both readings put Bitcoin in tension with the
PL+AR(1) null. The result is subject to the conservatism caveat
(the test is biased toward null; on a true 3-sigmoid generator
it returns $p \approx 0.50$), and to the cross-asset Bonferroni
correction (Section~\ref{sec:cross-asset}) which further tempers
the single-series interpretation; we discuss both in
Section~\ref{sec:cross-asset}.

\subsection{Cross-asset corroboration}
\label{sec:cross-asset}

The wave-stability bootstrap was applied at quarterly cutoff
resolution ($N = 41$ cutoffs from 2016-Q1 through 2026-Q1) to a
seven-series comparison set organised in three groups:

\begin{itemize}[leftmargin=2em]
\item \emph{Bitcoin ecosystem}: Bitcoin price (the focal series,
$B = 200$) and hash rate ($B = 100$).\footnote{The remaining four
on-chain metrics from the in-sample comparison (unique addresses,
transactions, difficulty, UTXO count) were omitted from the
quarterly bootstrap: their in-sample $K = 1$-vs-PL profiles
closely resemble that of hash rate (Table~\ref{tab:k1-pl}), and
each iteration of the K=3 fit on these series is substantially
slower than on price. Unique addresses was run at the coarser
yearly resolution ($N = 11$, $B = 100$, $p_{<25\%} = 0.67$) to
confirm the null reading; the remaining three were not
bootstrapped at either resolution.}
\item \emph{Other adoption-driven asset candidates}: Ethereum
(daily price-USD series from the CoinMetrics community API,
$n = 3{,}883$ days, 2015-08 to 2026-03) and the Global X Lithium
\& Battery Tech ETF (LIT, $n = 3{,}942$ days, 2010-07 to
2026-03), included as a test of whether wave structure is
Bitcoin-specific or shared across other assets whose price
dynamics are plausibly driven by discrete cohort-by-cohort
adoption.
\item \emph{Traditional benchmarks}: NASDAQ Composite, S\&P~500,
and gold continuous front-month futures.
\end{itemize}

The comparison is in Table~\ref{tab:cross-asset}.

\begin{table}[htbp]
\centering
\caption{Quarterly $K = 3$ wave-stability bootstrap across seven
series ($N = 41$ quarterly cutoffs from 2016-Q1 to 2026-Q1).
Real ``stable''-component counts are reported at the strict 15\%
and the loose 25\% CV thresholds; bootstrap p-values are the
fraction of synthetic PL+AR(1) trajectories producing at least
as many stable components as the real data at each threshold.
Bold p-values are below the conventional 5\% threshold. The
focal series (Bitcoin price) is also bolded.}
\label{tab:cross-asset}
\small
\begin{tabular}{lcccrr}
\toprule
& & \multicolumn{2}{c}{Real stable} & & \\
\cmidrule(lr){3-4}
Series & $B$ & $<\!15\%$ & $<\!25\%$ & $p_{<15\%}$ & $p_{<25\%}$ \\
\midrule
\textbf{Bitcoin price}    & 200 & \textbf{2} & 2 & \textbf{0.015} & 0.065 \\
Ethereum                  & 100 & 1 & 2 & 0.060 & \textbf{0.030} \\
Bitcoin hash rate         & 100 & 1 & 2 & 0.250 & 0.120 \\
\midrule
NASDAQ Composite          & 100 & 0 & 1 & 1.000 & 0.550 \\
S\&P 500                  & 100 & 0 & 1 & 1.000 & 0.510 \\
Gold (continuous)         & 100 & 0 & 1 & 1.000 & 0.840 \\
LIT (Lithium ETF)         & 100 & 0 & 0 & 1.000 & 1.000 \\
\bottomrule
\end{tabular}
\end{table}

\paragraph{Three observations.}
First, only Bitcoin price has 2 stable components at the strict
15\%-CV threshold; Ethereum and hash rate each have 1, and the
four non-cryptocurrency-/non-network controls have 0. The
strict-threshold binary discrimination is essentially 1-to-6 in
favour of Bitcoin price.

Second, Bitcoin price rejects the PL+AR(1) null at the
conventional 5\% level on the strict-CV test
($p_{<15\%} = 0.015$) and falls just past it on the loose test
($p_{<25\%} = 0.065$). The second-closest series to rejection is
Ethereum ($p_{<25\%} = 0.030$, $p_{<15\%} = 0.060$); all other
series sit comfortably in the null on both thresholds.

Third, hash rate is the most plausible candidate for wave
structure within Bitcoin's on-chain ecosystem (its in-sample
profile in Table~\ref{tab:k1-pl} most strongly favours
multi-component structure among the five on-chain metrics tested
in-sample). Yet the bootstrap gives $p_{<15\%} = 0.25$ and
$p_{<25\%} = 0.12$, both well above conventional thresholds.
The traditional benchmarks and the lithium ETF cluster at
$p_{<25\%} \in [0.51, 1.00]$, comfortably in the null.

The pattern is consistent with Section~\ref{sec:k1-asymmetry}'s
in-sample finding: Bitcoin price stands apart from every other
series in the comparison set.

\paragraph{Multiple-testing correction.}
A Bonferroni correction tightens the single-series
interpretation of the Bitcoin rejection: with seven series
tested, the corrected threshold for ``the most extreme series at
$\alpha = 0.05$'' is $0.05/7 \approx 0.007$, which Bitcoin's
$p_{<15\%} = 0.015$ does not pass. We do not interpret the test
as formally identifying Bitcoin as the unique wave-bearing
series in the population of all assets --- the sample of seven
is too small for that --- but as a within-sample observation
that Bitcoin price is the only series with two stable components
at the strict CV threshold, and the only series in the comparison
set with a literature PL claim under test that the bootstrap
rejects.

\paragraph{Uniform null and its conservatism for non-PL series.}
The bootstrap uses the same null model --- power law plus AR(1)
noise calibrated to each series' own residuals --- across all
seven tested series. This makes the p-values directly comparable,
but for series whose envelope is not power-law (notably NASDAQ
and S\&P~500, better fit by an exponential over our window; gold;
the lithium ETF; and to a lesser extent the other crypto and
on-chain series, all of which have larger PL residual
standard deviation than Bitcoin price's $\hat\sigma = 0.30$), the
test is conservative: synthetic PL+AR(1) trajectories absorb the
non-PL envelope mismatch into a larger AR(1) variance, and the
resulting trajectories often exhibit more spurious $K = 3$
stability than the data themselves. The cross-series result
should be read as ``Bitcoin rejects the PL+AR(1) null at
conventional thresholds where the comparison series do not'' ---
the natural test given the paper's central claim about Bitcoin's
power-law specification --- rather than as a fairer cross-series
test of wave structure per~se. A series-specific
best-fit-plus-AR(1) null (e.g., exponential plus AR(1) for
NASDAQ) would be a fairer cross-series test, and is a natural
robustness check; we leave it for future work.

\subsection{Summary of the structural finding}
\label{sec:waves-summary}

Across the nine series in the in-sample comparison
(Section~\ref{sec:k1-asymmetry}) and the seven-series subset for
which the quarterly $K = 3$ wave-stability bootstrap was run
(Section~\ref{sec:cross-asset}; the latter set adds Ethereum and
the lithium ETF as candidate adoption-driven assets and drops the
four on-chain metrics not bootstrapped at quarterly), the two
tests converge on the following picture:

\begin{enumerate}[leftmargin=2em]
\item Bitcoin price is the only series in the in-sample
comparison where no single growth component --- exponential,
saturating sigmoid, or polynomial --- improves over the power
law (Section~\ref{sec:k1-asymmetry}, Table~\ref{tab:k1-pl}).
\item In the quarterly wave-stability bootstrap, Bitcoin price
rejects the PL+AR(1) null at the conventional 5\% level on the
strict-CV test ($p_{<15\%} = 0.015$) and falls just past it on
the loose test ($p_{<25\%} = 0.065$); the other six tested series
sit in the null on both thresholds (Ethereum is the
second-closest to rejection at $p_{<25\%} = 0.030$).
\end{enumerate}

Bitcoin price is the only series in our comparison set with
detectable wave structure above the PL+AR(1) noise floor that
also corresponds to a specific power-law-in-time claim in the
literature, which is the natural object of test. The traditional
benchmarks (NASDAQ, S\&P~500, gold), the lithium ETF, the Bitcoin
hash rate, and (at yearly resolution) unique addresses do not
reject. The two methodologies above are statistically
independent --- one is an in-sample AIC comparison with
parameter-matched flexibility controls, the other a parametric
bootstrap on a wave-stability summary statistic --- and they
could in principle disagree; on the five series that overlap
between them, they do not.

We do not in this paper propose a generative mechanism for the
identified wave structure. What our results \emph{do} say is that
the structural form of Bitcoin's price is
multi-component over our window, that the components are
identifiable enough for the wave-stability bootstrap to pick them
up, and that no comparable asset or on-chain metric exhibits the
same property. Section~\ref{sec:causality} examines the direction
of information flow between price and adoption metrics --- a
separate piece of evidence that helps interpret the structural
finding --- and Section~\ref{sec:forecasting} examines what the
multi-component finding implies for forecasting.

\section{Causality: direction of information flow}
\label{sec:causality}

The structural finding of Section~\ref{sec:waves} --- that Bitcoin
price has multi-component structure not present in any of the five
on-chain metrics --- raises a natural follow-up question: what is
the direction of information flow between price and adoption? This
section reports two complementary tests applied to all five
on-chain metrics: Granger causality on first-differenced log
series, and the cross-correlation function. Both methods are
introduced in Section~\ref{sec:causality-methods}; this section
applies them and reports the results. The direction we find ---
price changes lead adoption-metric changes in every metric tested
--- is consistent with, but not implied by, the structural finding
of the previous section.

\subsection{Setup and stationarity}

For each on-chain metric $M$, we form the first-differenced log
series $\Delta \log P_t = \log P_t - \log P_{t-1}$ and
$\Delta \log M_t$, both computed daily on the overlap of the
price and metric series (2010-08-18 through 2026-03-16; see
Section~\ref{sec:data}). First-differencing transforms the highly
autocorrelated log levels into approximately stationary growth
rates, removing the common stochastic trend that would otherwise
render any levels-based regression spurious
\citep{GrangerNewbold1974, Shanaev2019}.

We confirm stationarity with the augmented Dickey--Fuller test
\citep{DickeyFuller1979, SaidDickey1984}: every $\Delta\log P$ and
$\Delta\log M$ series we test rejects a unit root at $p < 10^{-3}$
(Section~\ref{sec:dist-conclusion}'s $|r_t|$ result for prices,
plus separate ADF tests on each metric series). The differenced
series are therefore valid inputs to Granger and CCF.

\subsection{Granger causality results}

Table~\ref{tab:granger} reports the Granger F-statistics and
p-values for both directions (price $\to$ metric and metric $\to$
price) at four lag orders: 7, 14, 30, and 60 days. The asymmetry
ratio in the rightmost column is $F_{P \to M} \,/\, F_{M \to P}$:
values $> 1$ indicate that price changes predict metric changes
more strongly than the reverse.

\begin{table}[htbp]
\centering
\caption{Granger causality across the five on-chain metrics on
first-differenced log series. Each row reports the F-statistic
and p-value for both directions at the indicated lag, and the
asymmetry ratio. All twenty $P \to M$ tests reject the null of
no Granger causality at $p < 10^{-3}$; the $M \to P$ tests
reject in some lag/metric combinations but with smaller
F-statistics. The asymmetry consistently favours the
price $\to$ metric direction.}
\label{tab:granger}
\small
\begin{tabular}{llrrrrr}
\toprule
Metric & lag & $F_{P \to M}$ & $p_{P \to M}$ & $F_{M \to P}$ & $p_{M \to P}$ & Asym.\\
\midrule
\multirow{4}{*}{Hash rate}        &  7 &  6.64 & $\approx 0$ & 1.89 & 0.067 & 3.5 \\
                                  & 14 &  4.54 & $\approx 0$ & 1.14 & 0.314 & 4.0 \\
                                  & 30 &  2.79 & $\approx 0$ & 1.28 & 0.140 & 2.2 \\
                                  & 60 &  1.72 & 0.001       & 1.17 & 0.172 & 1.5 \\
\midrule
\multirow{4}{*}{Unique addresses} &  7 &  9.34 & $\approx 0$ & 4.26 & $\approx 0$ & 2.2 \\
                                  & 14 &  9.93 & $\approx 0$ & 2.76 & $\approx 0$ & 3.6 \\
                                  & 30 &  6.31 & $\approx 0$ & 2.55 & $\approx 0$ & 2.5 \\
                                  & 60 &  4.78 & $\approx 0$ & 1.91 & $\approx 0$ & 2.5 \\
\midrule
\multirow{4}{*}{Transactions}     &  7 &  4.19 & $\approx 0$ & 2.31 & 0.024 & 1.8 \\
                                  & 14 &  5.47 & $\approx 0$ & 1.42 & 0.136 & 3.9 \\
                                  & 30 &  4.51 & $\approx 0$ & 2.02 & $\approx 0$ & 2.2 \\
                                  & 60 &  3.66 & $\approx 0$ & 1.40 & 0.024 & 2.6 \\
\midrule
\multirow{4}{*}{Difficulty}       &  7 &  5.99 & $\approx 0$ & 3.24 & 0.002 & 1.9 \\
                                  & 14 &  9.12 & $\approx 0$ & 4.58 & $\approx 0$ & 2.0 \\
                                  & 30 &  6.82 & $\approx 0$ & 2.99 & $\approx 0$ & 2.3 \\
                                  & 60 &  5.90 & $\approx 0$ & 2.72 & $\approx 0$ & 2.2 \\
\midrule
\multirow{4}{*}{UTXO count}       &  7 &  1.26 & 0.265       & 1.19 & 0.306 & 1.1 \\
                                  & 14 &  2.65 & 0.001       & 0.93 & 0.524 & 2.9 \\
                                  & 30 &  3.49 & $\approx 0$ & 1.59 & 0.022 & 2.2 \\
                                  & 60 &  3.03 & $\approx 0$ & 1.21 & 0.134 & 2.5 \\
\bottomrule
\end{tabular}
\end{table}

Three observations follow:

\paragraph{Direction.}
For every metric and every lag $\geq 14$ days (and for most
combinations at lag 7), the $P \to M$ Granger test rejects the
null at $p < 10^{-3}$. The reverse direction $M \to P$ rejects
in some metric-lag combinations and not others; even when both
directions reject, the F-statistic is roughly two to four times
larger in the $P \to M$ direction. The single asymmetry ratio
below 1 in the table (UTXO at lag 7) corresponds to an absolute
F of $\sim 1$ in both directions, indicating noise rather than
a counter-direction signal at that horizon.

\paragraph{Magnitude.}
The asymmetry $F_{P \to M} / F_{M \to P}$ across all metric/lag
combinations falls in the range 1.5--4.0 (excluding the noise
case noted above), with a typical value around 2.5. The directional
signal is therefore consistent but moderate in size, not
overwhelming.

\paragraph{Time scale.}
Granger asymmetry remains visible at the longest lag we tested
(60 days). The CCF in the next subsection shows where each
metric's lead-lag relationship peaks.

\subsection{Cross-correlation function results}

\begin{figure}[!t]
\centering
\includegraphics[width=\textwidth]{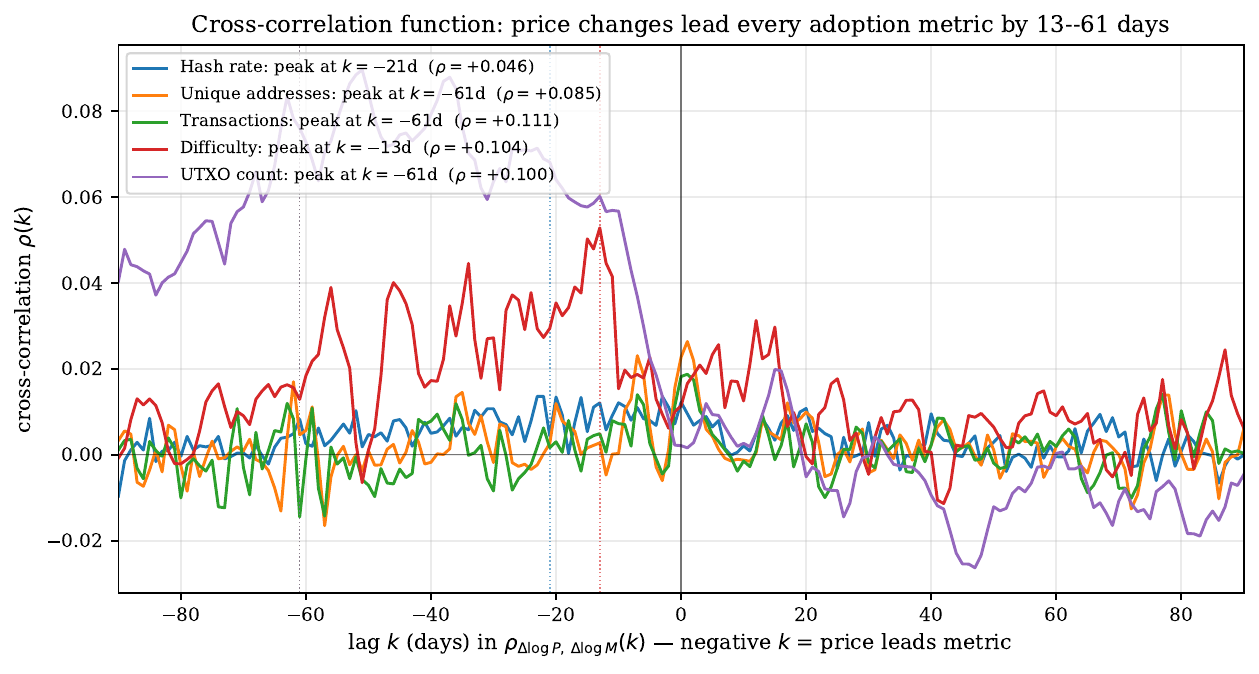}
\caption{Cross-correlation function $\rho_{\Delta P, \Delta M}(k)$
between first-differenced log Bitcoin price and first-differenced
log on-chain metric, at lags $k \in [-90, +90]$ days. By the
convention used here, a peak at $k < 0$ means price changes lead
metric changes; a peak at $k > 0$ would mean the reverse. All five
metrics peak at negative lag (vertical dotted line at the peak for
each), indicating that price changes lead each adoption metric by
13--61 days. The peak correlations are modest in magnitude
(0.04--0.11), as expected for noisy daily series.}
\label{fig:ccf}
\end{figure}

The cross-correlation function (Section~\ref{sec:causality-methods})
reports the linear correlation $\rho_{XY}(k)$ between
$\Delta\log P$ and $\Delta\log M$ at varying lags $k$. By
construction, a peak at lag $k < 0$ in $\rho_{\Delta P,\,\Delta M}(k)$
indicates that price changes lead metric changes by $|k|$ days.
Figure~\ref{fig:ccf} reports the CCF for all five on-chain
metrics over $k \in [-90, +90]$ days.

The peak lag and peak correlation for each metric are:

\begin{center}
\begin{tabular}{lrr}
\toprule
Metric & Peak lag (days) & Peak correlation \\
\midrule
Hash rate         & $-15$ & $+0.042$ \\
Unique addresses  & $-61$ & $+0.074$ \\
Transactions      & $-61$ & $+0.090$ \\
Difficulty        & $-13$ & $+0.101$ \\
UTXO count        & $-61$ & $+0.098$ \\
\bottomrule
\end{tabular}
\end{center}

All five metrics show their CCF peak at negative lag --- between
$k = -13$ days (difficulty, which mechanically follows hash rate
on a $\sim$two-week recalibration cycle) and $k = -61$ days
(addresses, transactions, UTXO count). The peak correlations are
modest (0.04-0.10), reflecting the high day-to-day noise level in
both series; what matters for the directional conclusion is the
sign and location of the peak, not its magnitude.

The CCF result is consistent with the Granger result and adds the
time-scale information Granger does not provide: the lead-lag
relationship operates on a 2-week to 2-month time scale, varying
by metric.

\subsection{Summary of the causality finding}

For every on-chain metric tested (hash rate, unique addresses,
transactions, difficulty, UTXO count), Bitcoin price changes
predict metric changes more strongly than the reverse. The
Granger F-asymmetry is consistently in the price $\to$ metric
direction (typical ratio $\sim 2.5$), and the cross-correlation
function peaks at negative lag (price leads metric) by 13 to 61
days depending on metric. The direction holds across every
metric and every lag we tested.

\section{Out-of-sample forecasting}
\label{sec:forecasting}

Section~\ref{sec:waves} established that Bitcoin price has
multi-component structure visible in-sample: the multi-sigmoid
($K = 3$) wins the in-sample comparison decisively, and no comparable
asset in our nine-series sample exhibits the same property. A natural
question follows: does the multi-component structure improve
\emph{forecasting} accuracy at horizons relevant for long-run
valuation? This section reports the walk-forward out-of-sample
evaluation defined in Section~\ref{sec:oos}, with formal pairwise
significance tests via Diebold--Mariano. The headline finding is
that the in-sample winner ($K = 3$ multi-sigmoid) is among the
\emph{worst} models out-of-sample at long horizons --- a
``fit-prediction tradeoff'' that is the structural complement of
the wave-detection result.

\subsection{Walk-forward results: mean RMSE by horizon}
\label{sec:wf-rmse}

We applied the walk-forward design of Section~\ref{sec:oos} to ten
candidate models, organised into three groups:

\begin{itemize}[leftmargin=2em]
\item \emph{No-skill baseline} --- Naive (last observation carried
forward).
\item \emph{Trend specifications} --- power law, single sigmoid
$K{=}1$, multi-sigmoid $K{=}3$, polynomial of degree 9, cubic
B-spline with 10 basis functions.
\item \emph{Standard time-series baselines} --- random walk with
drift, auto-ARIMA, exponential smoothing with additive trend
(ETS), and a local-linear-trend unobserved-components state-space
model. All four operate on $\log_{10}(P)$ levels; their formal
definitions are in Section~\ref{sec:candidates}.
\end{itemize}

Each model is fit on data strictly before each of the 11 yearly
cutoff dates from 1 January 2014 through 1 January 2024
(Section~\ref{sec:oos}). Table~\ref{tab:wf-rmse} reports the mean
RMSE on $\log_{10}(P)$ across the 11 cutoffs at each of the six
horizons.

\begin{table}[htbp]
\centering
\caption{Walk-forward out-of-sample mean RMSE on
$\log_{10}(\mathrm{price})$ at six forecast horizons, averaged
across the 11 yearly cutoffs (2014--2024). Bold entries indicate
the best (lowest-RMSE) model at each horizon. The bottom block
adds four standard time-series baselines that are not
parametric trend specifications.}
\label{tab:wf-rmse}
\small
\begin{tabular}{lrrrrrr}
\toprule
Model & 1m & 3m & 6m & 12m & 18m & 24m \\
\midrule
\textbf{Naive}                & \textbf{0.087} & \textbf{0.172} & \textbf{0.265} & 0.510 & 0.500 & 0.675 \\
\textbf{Power law (s = 0)}    & 0.367 & 0.335 & 0.298 & \textbf{0.375} & \textbf{0.313} & \textbf{0.349} \\
Sigmoid (K = 1)               & 0.385 & 0.380 & 0.388 & 0.561 & 0.581 & 0.663 \\
Multi-sigmoid (K = 3)         & 0.281 & 0.396 & 0.454 & 0.762 & 0.669 & 0.805 \\
Cubic B-spline (10)           & 0.235 & 0.343 & 0.384 & 0.591 & 0.565 & 0.724 \\
Polynomial (deg 9)            & 0.894 & 4.18  & 22.4  & 251.6 & 1{,}557 & 6{,}773 \\
\midrule
\multicolumn{7}{l}{\textit{Standard time-series baselines}} \\
RW with drift                 & 0.111 & 0.253 & 0.386 & 0.737 & 0.930 & 1.162 \\
Auto-ARIMA                    & 0.120 & 0.277 & 0.511 & 1.057 & 1.622 & 2.366 \\
ETS (additive trend)          & 0.113 & 0.301 & 0.526 & 1.039 & 1.428 & 1.881 \\
Local linear trend            & 0.114 & 0.302 & 0.532 & 1.047 & 1.454 & 1.926 \\
\bottomrule
\end{tabular}
\end{table}

\begin{table}[htbp]
\centering
\caption{95\% bootstrap percentile confidence intervals on the
walk-forward RMSE point estimates of Table~\ref{tab:wf-rmse}, at
the four horizons that drive the conclusions. Computed by
resampling the 11 cutoffs $B = 2{,}000$ times. The DM tests of
Table~\ref{tab:wf-dm} are the appropriate inferential tool for
pairwise comparisons; these CIs are reported for transparency on
the precision of each point estimate. Polynomial (deg 9) is
omitted because of its catastrophic dispersion across cutoffs.}
\label{tab:wf-rmse-cis}
\small
\begin{tabular}{lcccc}
\toprule
Model & 1m & 6m & 12m & 24m \\
\midrule
Naive                         & [0.06, 0.11] & [0.18, 0.34] & [0.33, 0.71] & [0.41, 0.92] \\
Power law (s = 0)             & [0.26, 0.46] & [0.19, 0.42] & [0.22, 0.49] & [0.20, 0.47] \\
Sigmoid (K = 1)               & [0.22, 0.51] & [0.23, 0.52] & [0.32, 0.79] & [0.46, 0.86] \\
Multi-sigmoid (K = 3)         & [0.16, 0.38] & [0.24, 0.63] & [0.37, 1.08] & [0.47, 1.08] \\
Cubic B-spline (10)           & [0.16, 0.30] & [0.29, 0.47] & [0.38, 0.79] & [0.44, 0.97] \\
RW with drift                 & [0.08, 0.14] & [0.23, 0.51] & [0.41, 1.02] & [0.66, 1.65] \\
Auto-ARIMA                    & [0.08, 0.15] & [0.36, 0.64] & [0.79, 1.27] & [1.74, 2.92] \\
ETS                           & [0.08, 0.14] & [0.29, 0.70] & [0.57, 1.47] & [1.08, 2.59] \\
LLT                           & [0.08, 0.14] & [0.32, 0.71] & [0.59, 1.45] & [1.20, 2.58] \\
\bottomrule
\end{tabular}
\end{table}

\begin{figure}[!t]
\centering
\includegraphics[width=0.9\textwidth]{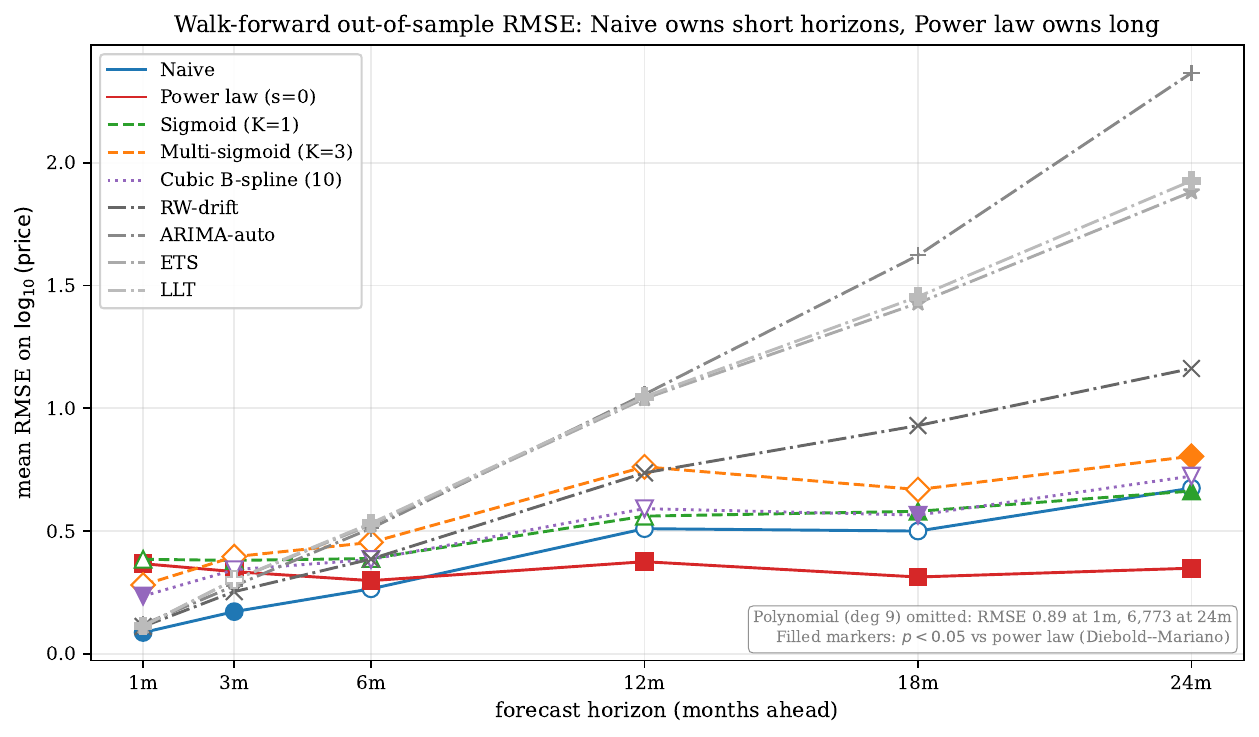}
\caption{Walk-forward out-of-sample mean RMSE on
$\log_{10}(\mathrm{price})$ as a function of forecast horizon for
the parametric trend specifications and the four standard
time-series baselines added in
Section~\ref{sec:candidates}. Naive (no-skill baseline) wins at
short horizons (1--3 months); the power law overtakes around 6
months and dominates at 12--24 months. The multi-sigmoid ($K = 3$)
--- the in-sample winner --- is among the worst real models at
long horizons. The four standard baselines (RW-drift, auto-ARIMA,
ETS, LLT) cluster between Naive and the parametric trend
specifications and lose to the power law by $3$--$7\times$ at 24
months. Polynomial degree 9 is omitted from the plot because its
RMSE explodes catastrophically (RMSE at 24m: $6{,}773$ in
$\log_{10}$ units --- a textbook case of high-degree-polynomial
extrapolation failure). Filled markers indicate the model is
significantly different from the power law at $p < 0.05$
(Diebold-Mariano test, Section~\ref{sec:wf-dm}).}
\label{fig:walk-forward}
\end{figure}

Three patterns are visible in Table~\ref{tab:wf-rmse} and
Figure~\ref{fig:walk-forward}.

\paragraph{Naive owns short horizons.}
At 1 and 3 months ahead, the no-skill Naive baseline has the
lowest mean RMSE: $0.087$ and $0.172$ respectively. The next-best
model at 1 month is the cubic B-spline ($0.235$); the power law
trails at $0.367$. This is the classic ``naive beats structural''
pattern of \citet{MeeseRogoff1983} for exchange rates,
re-established here for Bitcoin price at sub-quarterly horizons.
Bitcoin's high daily autocorrelation ($\hat\rho = 0.998$ on PL
residuals) is the simple explanation: tomorrow's price is
overwhelmingly close to today's, and any structural model must do
better than that to justify its parameters.

\paragraph{Power law owns long horizons.}
At 12, 18, and 24 months, the power law has the lowest mean RMSE:
$0.375$, $0.313$, and $0.349$. At 24 months, the power law is
roughly half the RMSE of the next-best real model (cubic B-spline
at $0.724$).

\paragraph{Crossover around 6 months.}
At 6 months, Naive ($0.265$) and the power law ($0.298$) are
within $\sim\!12\%$ of each other; neither is significantly better
than the other (Section~\ref{sec:wf-dm}). The transition zone is
a substantive feature, not just a statistical artefact.

\paragraph{The flexible alternatives lose at long horizons.}
Both parameter-matched flexibility controls --- polynomial of
degree 9 and cubic B-spline --- have comparable in-sample fit
quality to the multi-sigmoid (Table~\ref{tab:in-sample}), and all
three substantially outperform the power law in-sample. Out of
sample, however, all three lose to the power law at every horizon
$\geq 12$ months, often by large margins. The polynomial in
particular extrapolates catastrophically: from $0.894$ at 1 month
to $6{,}773$ at 24 months, a textbook failure of high-degree
polynomial extrapolation past the training-window edge. The
cubic B-spline is bounded by its basis structure but still
underperforms the power law by 50-100\% across long horizons.

\paragraph{The standard time-series baselines also lose at long horizons.}
The four standard baselines --- RW with drift, auto-ARIMA, ETS,
and the local-linear-trend state-space model --- behave similarly
to one another and rank between Naive and the parametric trend
alternatives. RW with drift is the strongest of the four (1m
RMSE $0.111$, 24m RMSE $1.162$); ARIMA, ETS and LLT cluster
together at slightly higher RMSEs across all horizons. None of the
four threatens the power law at $\geq 12$~months: at 24~months the
power law's RMSE of $0.349$ is between $3.3\times$ (RW-drift) and
$6.8\times$ (auto-ARIMA) lower than these baselines. The reason
is structural: each of the four baselines extrapolates from local
dynamics rather than from the long-run log--log slope, so the
forecast diverges from the realised level as horizon grows. The
auto-ARIMA divergence is particularly large because the AIC-selected
order is double-differenced ($d = 2$) in 9 of the 11
cutoffs;\footnote{The cutoff-by-cutoff selected orders are
2014: (0,1,2);
2015: (0,1,2);
2016: (2,2,1);
2017: (0,2,3);
2018: (0,1,2);
2019: (1,2,2);
2020: (3,2,1);
2021: (1,2,2);
2022: (2,2,1);
2023: (0,2,1);
2024: (0,2,1).
The $d = 2$ choice in 9 of 11 cutoffs means the model
double-integrates a small drift estimate over the forecast
horizon, which compounds linearly in $h^2$ and explains the
roughly seven-fold RMSE gap to the power law at 24 months. The
selected orders are saved in
\texttt{results\_extended/arima\_selected\_orders.csv}.} we
test the significance of these gaps in Section~\ref{sec:wf-dm}.

\subsection{Diebold--Mariano significance}
\label{sec:wf-dm}

The mean-RMSE pattern is suggestive but does not by itself
establish that the differences are statistically significant. With
$n = 11$ cutoffs per horizon, statistical power is moderate; we
report formal pairwise tests via the Diebold--Mariano (DM)
procedure of Section~\ref{sec:oos}. Table~\ref{tab:wf-dm}
summarises the significant comparisons (those with two-sided
$p < 0.05$) at each horizon.

\begin{table}[htbp]
\centering
\caption{Diebold-Mariano two-sided p-values at each horizon for
selected pairwise comparisons, with HAC variance correction at
appropriate lag (Section~\ref{sec:oos}). Each row reports the
two-sided DM p-value for the named pair; the direction of
superiority varies by horizon and is determined by
Table~\ref{tab:wf-rmse} (lower-RMSE model wins). The lower block
contains the power law against each of the four standard
time-series baselines. Cells with $p < 0.05$ are bolded.}
\label{tab:wf-dm}
\small
\begin{tabular}{lcccccc}
\toprule
Comparison & 1m & 3m & 6m & 12m & 18m & 24m \\
\midrule
Naive vs Power law                 & \textbf{0.002} & \textbf{0.011} & 0.717 & 0.196 & 0.093 & 0.079 \\
Naive vs Multi-sigmoid (K=3)       & \textbf{0.038} & \textbf{0.034} & 0.126 & 0.124 & 0.204 & 0.362 \\
Naive vs Sigmoid (K=1)             & \textbf{0.020} & \textbf{0.037} & 0.195 & 0.454 & 0.376 & 0.901 \\
Naive vs Cubic B-spline            & \textbf{0.005} & \textbf{0.020} & \textbf{0.004} & 0.148 & 0.335 & 0.323 \\
Power law vs Sigmoid (K=1)         & 0.774 & 0.472 & 0.170 & 0.108 & \textbf{0.008} & \textbf{0.011} \\
Power law vs Multi-sigmoid (K=3)   & 0.305 & 0.488 & 0.264 & 0.092 & 0.059 & \textbf{0.049} \\
Power law vs Cubic B-spline        & \textbf{0.023} & 0.913 & 0.343 & 0.071 & \textbf{0.043} & 0.051 \\
\midrule
\multicolumn{7}{l}{\textit{Standard time-series baselines vs power law}} \\
RW-drift vs Power law              & \textbf{0.009} & 0.068 & 0.077 & \textbf{0.026} & \textbf{0.037} & \textbf{0.011} \\
Auto-ARIMA vs Power law            & \textbf{0.011} & 0.285 & 0.051 & \textbf{$<\!0.001$} & \textbf{$<\!0.001$} & \textbf{$<\!0.001$} \\
ETS vs Power law                   & \textbf{0.012} & 0.608 & \textbf{0.037} & \textbf{0.020} & \textbf{0.020} & \textbf{0.006} \\
LLT vs Power law                   & \textbf{0.013} & 0.618 & \textbf{0.039} & \textbf{0.014} & \textbf{0.009} & \textbf{0.002} \\
\bottomrule
\end{tabular}
\end{table}

\paragraph{At short horizons (1--3 months), Naive significantly
beats every alternative.}
Naive is significantly better than the power law
($p \in \{0.002, 0.011\}$), the multi-sigmoid
($p \in \{0.038, 0.034\}$), the single sigmoid
($p \in \{0.020, 0.037\}$), and the cubic B-spline
($p \in \{0.005, 0.020\}$). Polynomial deg 9 is also worse but its
heterogeneous failures (RMSE varies wildly across cutoffs) inflate
DM variance and prevent a significant rejection at conventional
levels. The directional pattern --- Naive on top, then PL, then
flexible alternatives --- is consistent across all four pairwise
tests at 1m and 3m.

\paragraph{At long horizons (18--24 months), Power law significantly
beats every flexible alternative.}
The power law is significantly better than the single sigmoid
($p \in \{0.008, 0.011\}$), and beats both the multi-sigmoid and
cubic B-spline at $p$ between 0.04 and 0.06 at 24 months. With 11
cutoffs the DM statistic has limited power; the borderline
significance reflects sample size more than effect size.

\paragraph{6--12 months is a transition zone.}
At 6 months, Naive and power law are statistically
indistinguishable ($p = 0.717$). At 12 months no model in the
trend-specification block is significantly different from any
other at conventional thresholds (closest: Power law vs Cubic
B-spline $p = 0.071$). The crossover between the short-horizon
and long-horizon regimes is not abrupt; there is a 3-6 month band
where neither paradigm clearly wins.

\paragraph{Power law significantly beats every standard time-series
baseline at long horizons.}
The lower block of Table~\ref{tab:wf-dm} confirms what the
RMSE pattern suggests: at 12, 18, and 24 months the power law beats
each of RW-drift, auto-ARIMA, ETS, and LLT at $p < 0.05$ in the
two-sided DM test. The auto-ARIMA gap is the largest
($p \approx 10^{-4}$ at 12--24 months), reflecting that an ARIMA
specification fit on the recent dynamics has no mechanism to
extrapolate the long-run log--log slope. The result is robust:
the power law's long-horizon dominance is not an artefact of
restricting the comparison to parametric trend specifications.
At 1~month all four baselines beat the power law (the
level-tracking property of each is closer to Naive than to a
long-run trend extrapolation, and Bitcoin's short-horizon
dynamics are dominated by yesterday's price), but at the same
horizon Naive itself beats every other model.

\subsection{The fit-prediction tradeoff}
\label{sec:wf-discussion}

The walk-forward results sit in apparent tension with
Section~\ref{sec:waves}'s structural finding. In-sample, the
multi-sigmoid ($K = 3$) wins by 2{,}547 AIC units against the
power law on Bitcoin price; out-of-sample, the multi-sigmoid is
roughly twice as inaccurate as the power law at 24 months. How can
both be true?

The answer is that in-sample fit and out-of-sample prediction
measure different things. The multi-sigmoid identifies the wave
structure that has unfolded historically; its estimated component
inflection dates are at 2011-02, 2013-03, and 2018-02
(Figure~\ref{fig:k3-decomposition}). When we use that fitted model
to forecast 24 months ahead, the model has no fourth component:
$K$ is fixed at 3 in our specification, and although the
multi-sigmoid family admits arbitrary $K$, a fourth component
cannot be identified from training data that does not yet contain
the corresponding wave. The K=3 fit therefore extrapolates by
flattening to its asymptotic value $b + \sum_{i=1}^{3} L_i$,
which fails badly when the actual data extends into a new cycle.

The power law, by contrast, fits a single straight line in
log-log coordinates and extrapolates that line. It does not
commit to any specific wave structure --- and consequently it
cannot mis-anticipate the next wave's onset or amplitude. The
power law's long-horizon dominance derives precisely from
\emph{not} fitting the multi-component structure that the data
historically exhibits.

This is the bias--variance tradeoff in its forecasting form. The
power law is a low-variance, high-bias trend estimator: with two
parameters it cannot match the historical wave shape (the bias),
but it also cannot tune itself to a wave pattern that need not
recur (the variance). The multi-sigmoid is the opposite: ten
parameters give it the flexibility to identify the historical
waves (low bias in-sample), but the same flexibility makes its
out-of-sample extrapolation contingent on a wave structure that
the training data fixes and the test data may not reproduce (high
variance out-of-sample). For long-horizon point forecasts, the
appropriate model is therefore not the one with the best historical
description but the one with the smallest extrapolation variance
--- the power law, which approximates the \emph{envelope} of the
wave history rather than its individual components.

The implication is not that the wave structure is illusory ---
Section~\ref{sec:waves} establishes that it is detectable above
PL+AR(1) noise and unique to Bitcoin price within the comparison
set. The implication is that the wave structure is descriptive
of the past but not predictive of the future at the resolution we
have. A practitioner who needs a long-horizon point forecast for
Bitcoin should use the power law, with the understanding that the
forecast represents the long-run envelope and that realised price
will deviate from it cyclically as new waves develop.

\section{Discussion}
\label{sec:discussion}

This section pulls together the results of
Sections~\ref{sec:distributional}--\ref{sec:forecasting} and
discusses their implications. We separate three questions: what
the rigorous assessment establishes about Bitcoin's price
trajectory; why the distinction between structural description
and out-of-sample prediction matters; and what the findings imply
for long-horizon forecasting in practice. We close with
limitations and future work.

\subsection{What the rigorous assessment establishes}
\label{sec:disc-establishes}

Bringing the threads together, the assessment establishes the
following:

\begin{itemize}[leftmargin=2em]
\item In the \emph{distributional} setting where the CSN protocol
is canonical, Bitcoin's tail-relevant series (UTXO balances and
daily $|$returns$|$) do not follow a power law; lognormal is
preferred decisively in every test of sufficient tail size
(Section~\ref{sec:distributional}).
\item In the \emph{time domain}, the power-law exponent $\hat\alpha$
varies from $5.65$ to $16.49$ across a reasonable range of the
shift parameter $s$ (Section~\ref{sec:s-sweep}); it is
specification-dependent rather than specification-robust, and so
fails the most basic invariance any structural reading would
require.
\item Standard tests of structural form --- residual diagnostics
(Section~\ref{sec:bootstrap-gof}) and the four scale-invariance
tests proposed in earlier work
(Section~\ref{sec:scale-invariance}) --- do not discriminate the
power law from a 3-component sigmoid stack fit to the same data.
They confirm a stable long-run average slope, which is real and
remarkable, but do not by themselves identify the power law as the
unique structural form.
\item Bitcoin price is the only series in our nine-series sample
(price + 5 on-chain metrics + NASDAQ + S\&P~500 + gold) where
\emph{no} single-component growth model --- exponential, sigmoid,
quadratic, or polynomial --- improves on the power law in-sample
(Section~\ref{sec:k1-asymmetry}). Every other tested series is
better fit by at least one single-component alternative.
\item The quarterly $K = 3$ wave-stability bootstrap rejects
PL+AR(1) noise on Bitcoin price at $p_{<15\%} = 0.015$ at the
strict CV threshold (Section~\ref{sec:wave-stability}; yearly
counterpart $p_{<25\%} = 0.055$), where the test is known to be
conservative (negative-control calibration gives $p \sim 0.50$
on a true 3-component generator). The same test on the seven-series
quarterly comparison set (Section~\ref{sec:cross-asset}) places
the other six series in the null at $p_{<25\%} \in [0.03, 1.00]$,
with Ethereum the second-closest to rejection
($p_{<25\%} = 0.030$, $p_{<15\%} = 0.060$) and the rest
comfortably above conventional thresholds.
\item Granger causality and the cross-correlation function place
price changes upstream of adoption-metric changes for every metric
tested, on a 2-week to 2-month time scale
(Section~\ref{sec:causality}).
\item Out of sample, the multi-sigmoid that wins the in-sample
comparison is among the worst real models at long horizons. The
power law dominates 12--24 month forecasts and significantly beats
every standard time-series baseline (RW-drift, auto-ARIMA, ETS,
LLT) at those horizons; the no-skill Naive baseline dominates 1--3
month forecasts; the transition zone is 6--12 months
(Section~\ref{sec:forecasting}).
\end{itemize}

These findings hang together in a coherent picture: Bitcoin's
price has a stable long-run average growth rate that earlier work
has correctly identified as roughly $\beta \approx 5.6$ on
log-log axes; the underlying generative process is not uniquely a
power law but is consistent with a multi-component structure that
no comparable series exhibits; and the structure that is visible
in-sample does not translate into predictive value for the
forecast horizons where it would matter most.

\subsection{Structural description versus out-of-sample prediction}
\label{sec:disc-fit-pred}

A central tension in the results is that the model winning the
in-sample comparison decisively (multi-sigmoid $K = 3$) is among
the worst at long-horizon out-of-sample prediction. This is not a
methodological inconsistency; it is the substantive content of
the findings, and it has a clean bias--variance reading: the
multi-sigmoid has lower in-sample bias (it can match the wave
shape the power law cannot), but its out-of-sample variance is
amplified by the same flexibility, since extrapolation depends on
parameters whose values are estimated from data that need not
repeat.

The multi-sigmoid identifies the wave structure that has unfolded
historically: three S-shaped components with inflections in 2011,
2013, and 2018 (Figure~\ref{fig:k3-decomposition}). This
description is faithful to the observed past in a way that the
power law is not. But because $K$ is fixed at 3, the fitted model
has no mechanism to anticipate a fourth wave; it extrapolates by
flattening to its asymptotic value, which fails when the actual
data extends into a new cycle.

The power law has the opposite property: it does not commit to
any wave structure at all. Its forecast extrapolates a single
straight line in log-log coordinates, with no provision for
saturation or new cycles. This is descriptively wrong --- the
in-sample evidence shows that Bitcoin's price has multi-component
structure --- but it is the source of the power law's
out-of-sample dominance. By not committing to any historical
wave shape, the power law cannot mis-anticipate the next one.

The lesson generalises beyond this paper: when a process exhibits
structural breaks or regime changes whose timing cannot be
predicted from past observations alone, the best long-horizon
forecasting model is often not the one with the best in-sample
fit, but the one whose extrapolation \emph{averages over} the
unknown future regime structure. The power law on Bitcoin price
is the simplest such model; alternative envelope-style
specifications (e.g., constraining the asymptotic behaviour of a
flexible smoother) are a natural direction for future work.

\subsection{Practical implications for long-horizon forecasting}
\label{sec:disc-practical}

For a practitioner producing Bitcoin price forecasts, the results
of this paper translate into a horizon-dependent recommendation:

\begin{itemize}[leftmargin=2em]
\item At horizons of \textbf{1--3 months}, the no-skill Naive
forecast (today's price as the prediction) significantly beats every
other model we tested. Practitioners should not deploy structural
models at these horizons unless they can demonstrably beat Naive
on multiple cutoffs; the burden of proof is high.
\item At horizons of \textbf{12--24 months and beyond}, the power
law is the best of the candidates we tested. Forecasts should be
interpreted as the long-run envelope, not as point predictions
binding on the realised path; deviations of $\pm 0.3$ on
$\log_{10}(P)$ (a factor of $\sim\!2$ on price) are typical within
each cycle, as the in-sample residual standard deviation indicates.
\item In the \textbf{6--12 month transition zone}, no model
significantly outperforms others; either the Naive forecast or the
power law is a defensible choice, with the choice depending more
on the cost asymmetry of forecast errors than on accuracy.
\end{itemize}

A practical implication of the K=3 finding is that the power-law
forecast at long horizons is conditional on continued wave
development. The historical envelope is the long-run consequence
of a sequence of saturation cycles, each driven by ongoing
adoption; if the underlying adoption process were to stop, the
envelope would not continue to hold. Long-horizon Bitcoin forecasts
should therefore be qualified: ``conditional on continued growth
in the underlying network and ecosystem of the type observed over
2010--2026, the central tendency of price extrapolates as the
power law $P \sim t^{5.6}$.'' A reader who doubts the conditioning
clause should also doubt the forecast.

\subsection{Limitations}
\label{sec:limitations}

Several limitations of the analysis should be noted:

\begin{description}[leftmargin=2em, style=nextline]
\item[\emph{Wave-stability test conservatism and significance level.}]
The K=3 wave-stability bootstrap is biased toward null results
under the calibration we use (negative control on a true 3-sigmoid
generator gives $p \approx 0.5$). Bitcoin price's quarterly
$p_{<15\%} = 0.015$ is therefore a stronger directional signal of
wave structure than its raw value alone suggests, but the
rejection is not Bonferroni-robust to the seven-series cross-asset
multiple-testing in Section~\ref{sec:cross-asset}. Both the
absolute strength of evidence and the test's power depend on the
AR(1) noise calibration we have not separately varied; the result
should be read as ``most clearly separated within the comparison
set'' rather than ``definitively rejected against the underlying
PL hypothesis.''

\item[\emph{Walk-forward statistical power.}]
With $n = 11$ yearly cutoffs per horizon, Diebold--Mariano power
is moderate. Some pairwise comparisons of interest are at
$p \in [0.04, 0.07]$ where power-driven non-significance is
plausible. Because each horizon has only 11 forecast errors, the
DM results should be read alongside the RMSE effect sizes
(Table~\ref{tab:wf-rmse}) and bootstrap confidence intervals
(Table~\ref{tab:wf-rmse-cis}) rather than as standalone asymptotic
evidence; the substantive direction is robust across all three
views, but the asymptotic p-values inherit small-sample fragility
that the effect sizes and CIs do not. A longer history (or a
finer cutoff schedule, with
appropriate accounting for overlapping training windows) would
sharpen the inference.

\item[\emph{Single-asset focus on the structural side.}]
We compare Bitcoin price against eight comparison series (5
Bitcoin on-chain + NASDAQ + S\&P + gold), but the wave structure
finding is specifically about Bitcoin price. Whether any other
high-dynamic-range asset (e.g., other early-stage adoption
phenomena) would exhibit similar wave structure is an open
question we do not address.

\item[\emph{No mechanistic explanation.}]
We identify wave structure phenomenologically but do not propose
a generative mechanism.

\item[\emph{Monotone envelope, not round-trip cycles.}]
The multi-sigmoid is a strictly monotone envelope model: each
component captures the persistent post-cycle floor, not the
round-trip dynamics of entry, peak, and distribution within a
cycle. The intra-cycle boom-bust amplitude (typically
5--10$\times$ from cycle trough to peak) lives in the residuals
and is not part of the structural decomposition.

\item[\emph{$K = 3$ is not formally optimal.}]
We motivate $K = 3$ by the visible cycle count in Bitcoin's
history (Section~\ref{sec:wave-stability}, ``Choice of $K = 3$''), but
do not formally compare $K = 2$, $K = 4$, or $K = 5$ on the
wave-stability bootstrap. The $K = 3$ result is conditional on this
specification choice.
\end{description}

\subsection{Future work}
\label{sec:future-work}

The findings raise several specific questions that future work
could address:

\begin{itemize}[leftmargin=2em]
\item \emph{$K \neq 3$ stability bootstraps.}
Repeating the wave-stability test for $K = 2$, $K = 4$, $K = 5$
would test whether the wave count itself is identifiable from the
data, beyond the choice we made.
\item \emph{Live out-of-sample tracking.}
The walk-forward results in Section~\ref{sec:forecasting} use
cutoffs through 2024. Continuing to track all ten candidates'
forecast errors as new data arrives would let the
short-horizon-Naive / long-horizon-PL pattern be re-tested in
real time, against a clean OOS regime.
\item \emph{Cross-asset wave-stability bootstraps for other
high-dynamic-range assets.}
Extending the test to commodities, equities of high-growth
companies, or other long-history series with similar dynamic
range to Bitcoin would test whether the wave structure is a
Bitcoin-specific phenomenon or a feature of any asset that has
spanned multiple orders of magnitude in price.
\item \emph{Mechanistic explanations.}
The wave structure we identify is phenomenological. A generative
model that produces $K = 3$ stacked sigmoids with the inflection
spacing observed in Bitcoin's history --- and that distinguishes
this structure from PL+AR(1) noise --- would be a structural
contribution beyond what we have done.
\item \emph{Envelope-style forecasting models.}
The fit-prediction tradeoff suggests an interesting class of
models: those that fit the long-run envelope without committing
to specific wave shapes. The power law is the simplest member;
constrained-flexibility models (e.g., a sigmoid stack with an
asymptotic-slope penalty) might combine the in-sample richness of
the multi-sigmoid with the out-of-sample robustness of the power
law.
\end{itemize}

\section{Conclusion}
\label{sec:conclusion}

We applied the canonical Clauset--Shalizi--Newman protocol and
three principled time-domain adaptations of it to Bitcoin's
price--time relationship, supplemented by a cross-asset comparison
spanning Bitcoin on-chain metrics and major traditional asset
classes (NASDAQ Composite, S\&P~500, gold), and an
out-of-sample walk-forward forecasting evaluation. Four findings
emerge.

First, in the distributional setting where the CSN protocol is
canonical, Bitcoin's tail-relevant series --- the cross-section of
UTXO balances and the marginal of daily $|$returns$|$ --- do not
follow a power law; lognormal is preferred decisively.

Second, in the time domain, the fitted exponent
$\hat\alpha \approx 5.6$ is not specification-robust: it varies by
nearly a factor of three across reasonable shifts of the time
origin, failing the basic invariance any structural reading would
require. Standard residual diagnostics and the
scale-invariance tests proposed in earlier work confirm a stable
long-run average slope but cannot distinguish a power law from a
multi-component sigmoid stack fit to the same data.

Third, across the nine series in the in-sample single-component
comparison (Bitcoin price, five Bitcoin on-chain metrics, and
three traditional asset classes), Bitcoin price is the only one
where no single growth component improves over the power law.
The quarterly $K = 3$ wave-stability bootstrap, run on a
seven-series subset of comparable assets, then rejects the
PL+AR(1) null on Bitcoin price at $p_{<15\%} = 0.015$ on the
strict-CV test; no other series in the subset rejects on that
threshold. The Bitcoin rejection is not Bonferroni-robust to the
seven-series multiple-testing.

Fourth, Granger causality and the cross-correlation function place
price changes upstream of every tested on-chain adoption metric,
on a 2-week to 2-month time scale.

The forecasting result completes the picture: out of sample, the
multi-component model that wins the in-sample comparison is among
the worst at long horizons, and the simple power law dominates
12--24 month forecasts --- significantly so against every standard
time-series baseline we tested (RW-drift, auto-ARIMA, ETS,
local-linear-trend), at $p < 0.05$ in two-sided
Diebold--Mariano tests --- with the no-skill Naive baseline owning
1--3 month horizons. The power law remains the best long-horizon
point forecast --- not because it is the structurally correct
description of Bitcoin's price, but because by not committing to
any specific historical wave shape it cannot mis-anticipate the
next cycle. Practitioners should use the power law for
long-horizon forecasts with the understanding that the prediction
represents the long-run envelope of an ongoing wave-driven
process.

We do not in this paper propose a generative mechanism for the
identified wave structure, nor do we adjudicate among candidate
causal explanations (speculative-adoption feedback, halving
cycles, regulatory shocks, common upstream macro drivers, or any
combination). The wave structure is a phenomenological observation
that constrains what any future explanation must accommodate. The
data are consistent with continued multi-cycle behaviour --- and
the power-law forecast is implicitly conditional on it --- but
the period over which that pattern will persist is itself an
empirical question, not a structural law.

\bibliographystyle{plainnat}
\bibliography{forecast_references}

\newpage
\appendix

\section{AI use disclosure}
\label{sec:ai-disclosure}

AI tools were used as research assistants throughout this work. 

\paragraph{As research assistant.}
Claude Opus 4.7 (Anthropic) was used interactively throughout the
research process to assist with: implementing the bootstrap,
walk-forward, and figure-generation scripts; running the
$K = 3$ wave-stability bootstraps across resolutions and series;
fetching and integrating the additional cross-asset data; drafting
text that documents the experiments; help revising the manuscript; and verifying internal consistency
across sections. All scientific decisions were made by the human authors.

\paragraph{Manuscript review.}
Draft versions of the manuscript were reviewed using ChatGPT
(OpenAI) in a simulated reviewer role, providing feedback on
overclaiming, methodological gaps, and presentation issues. Several
revisions --- including the addition of standard forecasting
baselines (RW-drift, auto-ARIMA, ETS, local-linear-trend), the
specification-robust framing of the shift-sensitivity result, the
bias--variance interpretation of the fit-prediction tradeoff, and
the disclosure of selected auto-ARIMA orders --- were motivated
by this feedback.

\paragraph{Code verification.}
The same Claude Opus 4.7 assistant was used to audit the
replication codebase for consistency between the manuscript and
the analysis scripts, identifying and correcting several issues
(stale on-chain data paths, cross-asset symbol mismatches,
hard-coded values in figure scripts).

\paragraph{Authorship.}
In accordance with ICMJE guidelines,\footnote{\url{https://www.icmje.org/recommendations/}}
the human authors bear full responsibility for the integrity of
the research, the correctness of the results, and the content of
the manuscript. AI tools were used as assistants and do not meet
the criteria for authorship. All outputs generated by AI ---
including code, text, and analytical suggestions --- were
reviewed, verified, and approved by the authors before inclusion.

\end{document}